\documentclass[12pt,preprint]{aastex}

\shorttitle{BZ Cam Wind}
\shortauthors{Honeycutt, Kafka \& Robertson}

\begin{document}


\title{Wind Variability in BZ Camelopardalis}


\author{R.K. Honeycutt\altaffilmark{1}, S. Kafka\altaffilmark{2}, 
J.W. Robertson\altaffilmark{3}}

\altaffiltext{1}{Astronomy Department, Indiana University, Swain Hall West,
 Bloomington, IN 47405. E-mail: honey@astro.indiana.edu}

\altaffiltext{2}{Dept. of Terrestrial Magnetism, Carnegie Inst. of Washington,
 5241 Broad Branch Road NW, Washington, DC 2001. E-mail: skafka@dtm.ciw.edu}

\altaffiltext{3}{Arkansas Tech University, Dept. of Physical Sciences,
 1701 N. Boulder Ave., Russellville, AR 72801-2222. E-mail: jrobertson@atu.edu}

\begin{abstract}

Sequences of spectra of the nova-like cataclysmic variable (CV) BZ Cam were acquired
on 9 nights in 2005-2006 in order to study the time development of episodes
of wind activity known to occur frequently in this star.  We confirm the
results of Ringwald \& Naylor (1998) that the P-Cygni absorption components of the
lines mostly evolve from higher expansion velocity to lower velocity as an
episode progresses.  We also commonly find blueshifted emission components
in the H$\alpha$ line profile, whose velocities and durations strongly suggest
that they are also due to the the wind.  Curiously, Ringwald \& Naylor reported
common occurences of {\em redshifted} H$\alpha$ emission components in their 
BZ Cam spectra.  We have attributed these emission components in 
H$\alpha$ to occasions when gas concentrations in the bipolar wind (both frontside
and backside) become manifested as emission lines as they move beyond the disk's 
outer edge.  We also suggest, based on changes in the P-Cygni profiles during
an episode, that the progression from larger to smaller expansion velocities is 
due the higher velocity portions of a wind concentration moving beyond the edge of the 
continuum light of the disk first, leaving a net redward shift of the remaining 
absorption profile.

We derive a new orbital ephemeris for BZ Cam, using the radial velocity of the
core of the HeI 5876\AA\ line, finding P = 0.15353(4).  Using this period the wind 
episodes in BZ Cam are found to be concentrated near inferior conjuction of the
emission line source.  This result helps confirm that the winds in nova-like CVs 
are often phase dependent, in spite of the puzzling implication that such winds 
lack axisymmetry.  

We argue that the radiation-driven wind in BZ Cam receives an initial boost by acting
on gas that has been lifted above the disk by the interaction of the accretion stream
with the disk, thereby imposing flickering time scales onto the wind
events, as well as leading to an orbital modulation of the wind due to the
non-axisymmetric nature of the stream/disk interaction.  Simultaneous photometry
and spectroscopy were acquired on 3 nights in order to test the possible
connection between flickering continuum light and the strength of the frontside
wind.  We found strong agreement on one night, some agreement on another,
and no agreement on the third.   We suggest that some flickering events lead to
only backside winds which will not have associated P~Cygni profiles.  

\end{abstract}


\section{Introduction}

This is the first of an anticipated short series of papers on wind features 
in the optical spectra of nova-like (NL) cataclysmic variables (CVs).  
CV winds have mostly been studied at UV wavelengths accessible 
only from space.  However, opportunities for UV spectroscopy from spacecraft 
have been restricted in recent years.  Furthermore the time variability of the 
winds, which is the primary topic of this investigation, is more easily studied 
from the ground, where more numerous and longer sequences of spectroscopic and 
photometric monitoring are practical.

CVs are semi-detached interacting binary stars, in which 
the Roche-lobe-filling secondary star loses mass through its inner Lagrangian 
point to the white dwarf primary, forming an accretion disk (Warner 1995).
The NL subset of CVs have higher mass transfer rates, resulting in
a disk that is too hot for dwarf nova outbursts.  In NL CVs seen at low orbital 
inclination, disk photons can be scattered in an expanding wind,
resulting in blueshifted absorption line components (P-Cygni profiles) in lines 
such as CIV 1548-52.   In high inclination (eclipsing) systems the line profile 
from the wind becomes purely broad emission near zero velocity.  The general
observational properties of CV winds have been summerized in Drew (1997)and in 
Froning (2005); wind models have been reviewed in Proga (2005).  Maximum observed 
blueshifted absorption velocities reach $\sim$5000 km sec$^{-1}$.  Estimated mass loss 
rates due to the wind vary considerably but values of 
$\dot{M}_{wind}\sim$10$^{-2}\dot{M}$
of the accreted mass are not unusual.  The loss of angular momentum via the
wind may be an important fraction of the $\dot{L}$ needed to drive the mass
transfer; see discussions in Cannizzo \& Pudritz (1988), Livio \& Pringle
(1994), King \& Kolb (1995), and Knigge \& Drew (1997).
The P~Cygni lines are formed in a rotating bi-polar wind, but the mechanism
for driving the wind remains somewhat uncertain.  The most widely-accepted 
driving force is resonance scattering of photons from the hot  
accretion disk boundary layer, and/or hot white dwarf.

Occasional P-Cygni profiles in the optical spectra of NL CVs have been known for 
some time, but recent work has shown line profiles due to winds to be 
relatively common in certain lines, at least for some stars.  Kafka and 
Honeycutt (2004) discussed the use of the optical HeI lines and H$\alpha$ to study 
the properties of disk winds in NL CVs.  
The orbital dependence and the secular time dependence are
two important wind properties whose study is greatly eased, or made
practical altogether, by the use of ground-based optical data.  Examples
of wind studies using these optical lines include Kafka et al. 2003 (Q Cyg), 
2004 (BZ Cam, Q Cyg, HR Del, DI Lac, BT Mon, AT Cnc), and 2009 (V592 Cas).  The
HeI triplets (2$^{3}$P $\rightarrow$ 3$^{3}$S and 2$^{3}$P 
$\rightarrow$ 3$^{3}$D for the 5876{\AA} and 7065{\AA} 
lines respectively) are particularly wind-sensitive, showing P~Cygni profiles 
when the wind is present, whereas  the singlet
He I line at $\lambda$6678 arises primarily from the disk.  Radiative transfer
effects in HeI have been studied by Almog and Netzer (1989) and by
Benjamin, Skillman \& Smits (2002).  These lines have found particular
application to the spectra of symbiotic stars (Proga, Mikolajewski \& Kenyon
1994; Silviero \& Munari 2003).

The strong metastability of the 2$^{3}$S level of
He I (which is the effective ground state of the triplets) likely alters
the recombination cascade in such a way as to make the triplets
more liable to absorption under the low density conditions and  
dilute radiation field of the wind, compared to the singlet at 6678\AA.   When P-Cygni 
profiles are present in the profiles of the HeI triplet lines, 
H$\alpha$ also often shows wind features. 
Among numerous NL CVs which we have examnined for wind line profiles in the optical, 
BZ Cam is among the most active, having wind features about
half the time, which is about the same rate found in optical spectroscopy
of BZ Cam by Ringwald \& Naylor (1998; hereafter RN98).  BZ Cam also has wind signatures 
in space-UV lines. (e.g. Hollis et al. 1992; Prinja et al. 2000). 

The rapid variability of the BZ Cam wind has received past attention,
including Griffith et al. (1995), Patterson et al.(1996), RN98, 
Prinja et al.(2000), and Greiner et al. (2001); episodic outflows have also
been described in other NLs (e.g. Prinja et al. 2000 for V603 Aql; Prinja et al.
2003 for RW Sex).  In general these studies describe wind events with lengths from
100 to 3000 sec, having blue-shifted absorption components reaching 
-3000 km s$^{-1}$, and little or no orbital dependence of the wind. 
RN98 report that the BZ Cam wind decelerates as an episode progresses, unlike the wind in 
OB stars.  This present study incorporates optical spectroscopy of BZ Cam acquired  
during 2005-2006..  This is a considerably
richer spectroscopic data set than earlier studies, allowing us to refine earlier 
results in the literature and also reveal additional wind phenonomena in BZ Cam.  
We also explore extensive long-term photometry of BZ Cam obtained 1990-2012.  

BZ Cam is at a distance of 830$\pm$160 pc (RN98).  The energy distribution in the
optical (Patterson et al. 1966) is quite blue, indicating a high $\dot{M}$
accretion system as is typical of most NL CVs which show winds.  The orbital
inclination is low, as judged by the lack of eclipses and the fact that the
cores of the emission lines are quite narrow ($\sim$12\AA\ FWHM).  RN98 estimate
an inclination of 12-40$\arcdeg$.

BZ Cam is unusual for being the only CV embedded in a 
nebula that is not related to a recorded nova explosion. It has been 
suggested that interactions of the BZ Cam wind with the interstellar medium 
produces the observed nebular bow shock (Hollis et al. 1992), but 
alternative origins for the nebula have also been proposed (e.g. Griffith, 
Fabian \& Sion 1995; Greiner et al. 2001).  

\section{Data Acquisition and Reductions}

\subsection{Photometry}

Table 1 is a log of the photometric observations.  Column 1 provides a reference
designation for each sequence,
column 2 is a UT or UT range, while column 3 gives similar information for the JD.  
Column 4 is the telescope, column 5 the exposure time in sec, column 6 the number 
of useable exposures, and column 7 the duration of the data stream. 
Column 8 indicates which (if any) spectral sequence was simultaneous 
with the photometry.  All of the photometry is in the V band.

Our long-term BZ Cam photometry, having typical spacings of days over intervals
of many years, consists of three sets 1990-1996, 2000-2005, and 
2007-2012, all using autonomous unattended telescopes in central Indiana.  The first two 
such sequences (P1 and P2) were obtained using a 0.41-m telescope (informally called 
RoboScope; Honeycutt et al. 1994 and references therein).  Flats and other detector 
calibration data were automatically acquired and applied each night, followed by aperture 
photometry and field identification, all using custom software (Honeycutt \& Turner 1992).  
Final photometric reductions were done
using the incomplete ensemble technique contained in Astrovar, which is a custom 
package based on the technique described in Honeycutt (1992), but with the 
addition of a graphical user interface.  The RoboScope Astrovar solution used
32 ensemble stars, and the zeropoint was established  to within 0.01 mag using 10 secondary 
standards from Henden \& Honeycutt (1995).  The third long-term sequence (P3) was 
acquired 2007-2012 using an unattended, autonomous 1.25-m telescope at the same site as
RoboScope.  The P3 data were reduced using a custom pipeline consisting of 
IRAF\footnote{IRAF is distributed by the National Optical Astronomy
  Observatories, which are operated by the Association of Universities for
  Research in Astronomy, Inc., under cooperative agreement with the National
  Science Foundation.} 
routines for detector calibrations, followed by the application of 
SExtractor\footnote{SEextractor is a source detection and photomery package
  described by Bertin and Arnouts 1996.  It is available from 
  http://terapix.iap.fr/soft/sextractor/.} for aperture photometry.  The light curves 
were then generated 
using Astrovar, employing a total of 24 ensemble comparison stars.  The zero point
was determined to within 0.01 mag using 11 secondary standards from Henden \& 
Honeycutt (1995).  Sequence P3 is missing data between 2009-Apr-25 and 2010-Oct-07 (the
full 2009-2010 winter observing season for BZ Cam) because of a detector failure.

Sequences P4, P5, and P6 are short-term photometry having typical spacings of
minutes over a single night each.  Data for P4 are from the Tenagra 
Observatory\footnote{http://www.tenagraobservatories.com/} 0.8-m telescope in southern 
Arizona and the data for Sequences P5 and P6 are from the 0.91-m 
WIYN\footnote{The WIYN Observatory is 
  a joint facility of the University of Wisconsin-Madison, Indiana University, 
  Yale University, and the National Optical Astronomy Observatory} telescope at Kitt 
Peak.  The purpose of these sequences was to provide photometry simultaneous with
spectroscopic sequences.  Both the Tenagra and WIYN 0.91-m exposures were reduced using 
differential photometry from C-Munipack\footnote{
  http://C-Munipack.sourceforge.net}, a PC (Windows)-based photometry package orginally 
developed by Hroch (1998) and maintained by David Motl.  This processing included detector 
calibrations as well as aperture photometry.  The differential photometry was placed on
the standard system using comparison star HH6 (V=14.00) from Henden \& Honeycutt
(1995).  Errors were assigned using the check star HH02 (V=13.38) 

Figure 1 shows our full light curve made up of all the sequences in Table 1.  
There are unfortunate gaps in the Indiana long-term coverage, especially 1996-2000.  
Compilation of averages of visual estimates in the AAVSO archives do not begin until 
2000-Dec and therefore do not help much with filling this gap.
During the small intervals of overlap the AAVSO visual estimates and the Indiana data 
agree reasonably well (to within $\sim$0.1 mag), except for a number of isolated faint 
AAVSO visual estimates at V = 14.2-15.6 between 2007-Mar and 2009-Apr.  There are
no similar faint data points in the relatively complete Indiana data over this same 
time interval, so it will be important to learn if such faint states actually occur in
BZ Cam.  We note that Garnavich \& Szkody (1988) found a 1928 low state of BZ Cam at
V=14.1, using the Harvard Plate Collection.  Also shown in Figure 1 are our straight line 
characterizations of the light curve from Greiner et al. (2001), whose data are from a 
variety of sources but mostly VSNET.  During the interval
of overlap from JD 2451780 to 2451870 the data in the Greiner et al. compilation are 
$\sim$0.35 mag brighter than the Indiana photometry.  During this same interval the 
Indiana photometry agrees with the AAVSO visual estimates to $\sim$0.1 mag.

Table 2 lists all the JDs, magnitudes, and errors (with respect to the ensemble) of the new
Indiana data.  The complete version of Table 2 is available only in electronic form.  
Errors (not shown in Figure 1 but included in Table 1) are for differential magnitudes with 
respect to the ensemble, and are mostly $\sim$0.02 mag.  

\subsection{Spectroscopy}

Our time-resolved spectroscopic data were obtained using a variety of
telescopes during the interval 2005-Oct to 2006-Sep.  The spectroscopic
sequences totaled 26 hours on 9 nights.  The wavelength
regions varied somewhat but always included the four lines HeI 5876\AA,
H$\alpha$, HeI 6678\AA, and HeI 7065\AA.

Table 3 is a journal of our spectroscopic runs.  Column 1 assigns a designation
to each spectral sequence for subsequent reference, while columns
2 and 3 give the UT date and the JD.  Column 4 is the telescope, column 5
is the exposure time in sec, and column 6 is the number of useable exposures
in the sequence.  Column 7 gives the duration of the sequence in hours, column
8 indicates if simulaneous photometry is available (see Table 1), and column 9 notes 
whether the sky was clear, mostly clear, or partly cloudy.  Figure 2 is an expanded 
portion of the BZ Cam light curve of Figure 1, with the times of our spectra marked.   

Spectral sets S1, S2, and S3 were obtained using the GoldCam slit spectrograph on the 
KPNO\footnote{Kitt Peak National Observatory is a division of the National 
  Optical Astronomy Observatory, which is operated by the Association of Universities for 
  Research in Astronomy, Inc., under cooperative agreement with the National Science
  Foundation} 
2.1-m telescope.  Grating 35 was used in first order, providing coverage 
$\sim$5400-7400\AA\  at $\sim$3\AA\  resolution.  
Spectral sets S4, S7, and S8 were obtained using the MOS/Hydra 
multiple object spectrograph on the WIYN\footnote{The WIYN Observatory is a joint 
  facility of the University of Wisconsin-Madison, Indiana University, Yale University, 
  and the National Optical Astronomy Observatory.}  
telescope.  Grating 600 was used in first order, 
providing coverage $\sim$5300-8200\AA.  The ``red'' 2$\arcsec$ fiber bundle was employed,
yielding $\sim$3\AA\ resolution; numerous other fibers were used for sky
subtraction.  Spectral sets S5 and S6 were obtained using the RC slit spectrograph 
on the Kitt Peak 4-m telescope.  Grating KPC-007 was used in first order, providing 
coverage $\sim$5600-7900\AA\ at $\sim$3\AA\  resolution.   
Spectral set S9 used the 6.5-m MMT\footnote{The MMT Obervatory is a joint facility 
 of the Smithsonian Institution and the University of Arizona} 
at Mt. Hopkins, Arizona.  The Blue Channel slit spectrograph was employed with a 
1200 line/mm grating, providing coverage $\sim$5800-7100\AA\ at a resolution of $\sim$1.5\AA.  

For all spectra the detector calibrations used standard 
IRAF\footnote{IRAF is distributed by the National Optical
  Astronomy Observatories, which are operated by the Association of
  Universities for Research in Astronomy, Inc., under cooperative
  agreement with the National Science Foundation.} 
procedures, and for spectral extractions and wavelength calibrations we used 
IRAF's onedspec/twodspec packages.  No spectrophotometric calibrations were
applied for any of the spectra, and the continua in the reduced spectra were nomalized 
to unity. 

\section{Qualitative Results and Comparisons with Earlier Work}
 
For NL CVs prone to wind activity, it has been found (Kafka \& Honeycutt 2004) that 
the HeI triplet lines at  5876\AA\ and 7065\AA\ often display wind 
features in their line profiles, while the HeI singlet at 6678\AA\  never shows wind 
features.  This behavior is fully confirmed in our BZ Cam spectroscopy.  Because 
HeI 6678 is uninteresting insofar as wind studies, and because the HeI 7065\AA\ line 
profile seems to simply be a weaker (and therefore noiser) version of the HeI 5876\AA\ 
profile, we have  concentrated our attention on just two lines: HeI 5876\AA\ and H$\alpha$.

The line profiles for the HeI 5876 line and for H$\alpha$ are displayed in two
different complementary ways.  In Figure 2 we show nested plots of the line profiles,
while in Figure 3 we show the same profiles as simulated trailed spectra with
the intensities coded as scales of grey.  The nested plots are best for revealing
relative strengths and shapes of wind features, while the trailed spectra are better 
for showing the time dependence of the line profiles.  The two presentations 
use the same data.  However, the trailed spectra have been rather strongly
contrast-enhanced to emphasize the time development of the wind episodes.   
We encourage
the reader to use both sets of figures as we describe the variablity in
the BZ Cam wind.  Due to clouds, the spectra in sequences S6, S7, and S8 are sometimes
quite noisy.  We nevertheless have included them in our analysis because the wind
features are still visible, and the information provided on duty cycles and time
dependences remain valuable to our study.

The P-Cygni profiles seen in Figures 2 and 3 occur in short bursts of activity, with
durations ranging from 7 to 90 min.  The mean is $\sim$45 min.  Because the
duty cycle of the episodes is near 50\%, some of the longer duration episodes 
probably are overlapping shorter episodes.  The mode of the distribution of episode 
durations ($\sim$35 min) may therefore be more meaningful.  These 
episode durations are consistent with that found by RN98
of $\sim$40 min for spectra on two nights in 1995 over 7.5 total hours.  It
is likely that our exposure times undersample some of the more rapid changes.
In fact, Prinja et al. (2000) reported wind variability times scales down to 100 
sec in HST UV spectra of BZ Cam.  Nevertheless, except at the start of the wind
episodes, there are no discontinuities
in the line profiles between adjacent spectra in Figures 2 and 3, leading us to conclude
that undersampling is not a serious problem for our data.

As seen in Figure 1, BZ Cam was fairly stable in brightness 1991-96, followed by
erratic variations in the range V=12-14 during 1997-2012, on time scales of weeks to 
months.  Although photometric coverage near the times of our spectra is spotty,
it appears from examination of Figure 1 that our spectra were acquired
when the system was near V = 13.3, which would be an intermediate (or
even intermediate/faint) state.  The fact that the BZ Cam wind is quite active
in our spectra is at variance with the conclusion of Greiner et al. (2001) that
the BZ Cam wind shows up only in the high state and is absent during the optical
low and intermediate states.  It may be that the BZ Cam wind varies on many time
scales, some of which have yet to be sampled.

As can be seen in Figure 3, the episodes of P~Cygni absorption seem to begin
abruptly and simultaneously in HeI 5876 and in H$\alpha$, and the evolution of the
profiles during an event are similar for two lines.  In most of the wind episodes 
the blueshifted absorption evolves towards slower expansion velocity as the event 
progresses.   Although they start at the same time, the blueshifted
absorption components of the line profiles typically persist for $\sim$2$\times$ longer
in HeI than in H$\alpha$.

RN98 reported a blueward linear increase in velocity for the first
6-8 min of a wind episode in BZ Cam, followed by a deceleration to near rest
in 30-40 min.  We see a rather wider range of behavior with no evidence for
an acceleration phase.  However, our results may be consistent RM98 when one 
recognizes that 
our spectral sequences would barely resolve a 6-8 min velocity increase to 
begin the wind event (RN98 had 60 s time resolution).  As best seen in Figure 2,
the blue absorption in our spectra typically begins as a broad, shallow feature,
which becomes deeper as the event progresses, as well as moving redward in mean
velocity as the bluer portions of the absorption disappear.  This happens in both
HeI and H$\alpha$ but is easier to see in HeI because the events last longer and
the absorption is often stronger.  Other changes sometimes occur in the shape of the
blueshifted absorption as the event progresses, such as development of sawtoothed 
profiles or multiple components (e.g. profiles 3-50, 3-51, 4-09, 4-10, 5-10, 5-12.)  
These more complicated profiles are nevertheless usually superimposed on the
general trend from broad shallow absorption to more narrow, deeper, less blueshifted
profiles as the event progresses.   The largest blueshifted absorption velocities
are $\sim$-2350 km s$^{-1}$, compared to maximum velocities near -3000 km s$^{-1}$ in 
RN98.

As best seen in Figure 3 the H$\alpha$ line profile often shows a blueshifted
emission component, which is never seen in HeI. 
The velocities of the blueshifted H$\alpha$ emission ranges up to 
-2200 km s$^{-1}$, similar to that of the blueshifted absorption in HeI 5876\AA\
and in H$\alpha$.  We do not see much systematic velocity 
evolution of the blueshifted H$\alpha$ emission during a wind event.  The velocities of
the emission in most of sequences appears random; however, there is some evolution
towards lower velocities in sequences S2 and S5.  The blueshifted H$\alpha$ emission
seems to often come after the appearance of blueshifted absorption in HeI and in 
H$\alpha$.  However, the time relationship of these events is difficult to pin down in our
data sets because the typical length of our spectral sequences not much longer
than the typical spacing between events, and because wind events sometimes overlap.
Because the time relationship of the various wind manifestions is important to
our interpretations, we will return to this point in later discussion. 

RN98 report occasional blueshifted H$\alpha$ emission in BZ Cam spectra, but the 
appearences were relatively brief, weak, and much less conspicious than in our 
BZ Cam spectra.  RN98 also 
saw frequent and conspicious {\em redshifted} H$\alpha$ emission in their BZ Cam
spectra, reaching up to +2400 km s$^{-1}$  (which can also be seen in the spectra of 
Patterson et al. 1996).  Curiously we do not see redshifted H$\alpha$ emission in {\em any} 
of our BZ Cam spectra reported here (though we did see redshifted H$\alpha$ emission
in a single BZ Cam spectrum acquired in 2004 (Kafka \& Honeycutt 2004)).  Both the redshifted 
and blueshifted H$\alpha$ emission features
are surely due to the wind, because they have velocities and time scales similar to the
P~Cygni absorption features.

The wind features seen in Sequence S9 (the MMT spectra) are different from the behaviors seen
in the other sequences.  In S9 the blue absorption in HeI is present nearly 
all the time, mostly at the same relatively low velocity.  There is a short wind episode
near the end of the Sequence S9 having the characteristic decline from high
to low expansion velocities.  However, in HeI this episode is weak and not well-defined.  The 
H$\alpha$ profile shows broad weak absorption at high negative velocities for the first 
2/3 of the sequence, followed by a brief wind episode at the end of the sequence, 
whose timing and velocities mirror that seen in HeI episode. There is occasional absorption
in the red wing of H$\alpha$, for which a physical explanation is lacking.   Furthermore 
that feature (all well as a similar absorption feature in the blue wing of HeI) seem to vary 
with a quasi-periodicity of $\sim$8-9 min (see Fig 3e).  (However, when the residual 
intensities in the relevant spectral windows were plotted vs. time, no meaningful 
quasi-periodicities were apparent.)  Sequence S9 has the 
most favorable combination of spectral resolution and S/N of our 9 sequences, 
so it is possible that these relatively weak behaviors are also present in the other 
sequences but are simply not apparent in spectra having lower S/N and/or lower spectral
resolution.  Because the S9 features are at variance with those in S1-S8, we considered
whether unknown data acquisition or data reduction problems could be responsible.
However, such features are lacking in the spectra of other CVs 
acquired on the same night, and we are inclined to regard the odd line profiles in 
the S9 spectra as real.  

\subsection{Orbital Ephemeris}

A reliable orbital ephermeris for BZ Cam is desired for two purposes:  1) to
correct the spectra for orbital motion before extracting velocities of
the wind features, and 2) to investigate any possible dependence of the
wind parameters on orbital phase.  We expected that our fairly extensive
spectroscopy (9 nights, well-distributed over a year, and having 4 emission 
lines) would allow determination of an accurate and reliable ephemeris.  To minimize
wind contributions
we measured the radial velocities of the central emission cores of H$\alpha$, 
HeI 5876\AA, HeI 6678\AA, and HeI 7065\AA\ in all our spectra.  (HeI 7065\AA\ was 
sometimes too faint for reliable measurement.)   However, preliminary periodogram 
analysis of these velocities 
proved discouraging.  Systematic radial velocity variations with typical
full amplitudes of $\sim$150 km s$^{-1}$ are present for most of the lines
for most of the runs in Table 1, all having characteristic  ``periods'' near
3.7 hr.  However, periodograms using the full year
of data seldom show significant peaks.  One exception is the periodogram for 
HeI 5876 which has a significant peak near 0.1535 days.  
This peak is consistent with the spectroscopic period of Lu and Hutchings (1985)
of 0.1535 d (using 49 spectra over 134 days) and with the Patterson et al. (1996) 
spectroscopic period of 0.1533(3) d (using 28 spectra over 5 successive nights).  
However, it is inconsistent with the Patterson et al. photometric period
of 0.153693(7).  (This photometric period used extensive photometry of BZ~Cam 
over $\sim$4 months in 1994-95, and nearby aliases were addressed by appealing
to the radial velocity data of Lu \& Hutchings (1985) as well as to the possible
spectroscopic periods in the Patterson et al. spectra.  Patterson et al. suggest that 
their photometric period might be affected by superhumps.)  Unfortunately the 0.1535d 
period lacks sufficient accuracy for use over our one year of 2005-2006 spectroscopy, 
and the photometric period of Patterson et al. (1996) does not phase our
radial velocity data at all.

Upon folding the He 5876 radial velocity curve on this initial period it
was obvious that much of the difficulty was due to a varying gamma velocity
from night to night.  To address this deficiency we "pre-whitened" the data
by adjusting the velocites for each night to each have an average of zero 
velocity.
This is an imperfect process because many of the 9 spectral sequences cover
less than a full orbit, causing the mean velocity to depend on the particular
phases that were covered.  Nevertheless this prewhitening dramatically 
increased the reliability of the detected period and resulted in a 
folded radial velocity curve of reasonably good quality (Figure 4).  The resulting
ephemeris for the times of gamma crossing from - to + velocity
for the HeI 5876 emission core is 

T$_{0}$ = 2453654.008(2) + 0.15353(4)*E.

This folded radial velocity curve has K = 84(5) km s$^{-1}$.  This new ephemeris
has sufficient accuracy to phase our spectroscopy over one year, and is the one we will use in 
this paper.  Assuming that HeI 5876 emission arises from the vicinity of the 
mass-gaining star (a reasonable assumption considering the strong wind profile seen
in this line), then phase zero corresponds to superior conjunction of the mass-losing
star.

Our new period also appears in a periodogram (the periodgrams are not shown) of 
the radial velocities of the H$\alpha$ emission core, at about the same accuracy 
and sharpness as in HeI 5876, but 
it is only the fourth strongest peak in the interval 0.1 to 0.2 days.  This
is in contrast to the periodograms of He I 5876, where this period is the
strongest in the interval 0.1-0.2 days, for both the raw and the prewhitened
radial velocities.  One of the reasons that the peak is missing or weak in some
lines can be seen in Figure 5.  In this figure we compare, as an example, 
the radial velocity curves for four spectral lines from Sequence S5.  We see that 
the velocity curves of HeI~5876 and HeI 6678 resemble one another, but that 
HeI 7065 and H$\alpha$ are nearly antiphased to that of $\lambda$5876 and 
$\lambda$6678.  The singlet HeI 6678 line seems to never display P-Cygni profiles, 
unlike either  H$\alpha$ or
the two triplet He I lines ($\lambda$5876 and $\lambda$7065); this holds true in both 
in BZ~Cam and in other CV winds (Kafka \& Honeycutt 2004 and refs therein).  
We might therefore expect the $\lambda$6678 radial velocity curve to be the least affected
by the wind and to be the ``odd man out'' among the lines, but that is not the case.
Spectral sequences other than S5 show similar discordant behaviors, but less pronounced.

Faced with such divergent behaviors among the radial velocity curves
of different lines at the same epoch, we choose to use the He I 5876 ephemeris
which (at least when prewhitened) does give a good period that is consistent
with earlier work.  We note that the three Balmer lines used in Lu \& Hutchings 
(1985) were also found to sometimes give differing velocities, and similar effects 
are present in the radial velocities of Patterson et al. 1996) data as well.
We suspect that the wind is subtly distorting the emission line profiles in BZ~Cam
even in the cores of the lines, and even when P~Cygni profiles are not apparent.
We conjecture that this distortion differs
from line-to-line, much as the P~Cygni profiles differ from line-to-line, giving
rise to discordant r.v. curves for different lines.  For poorly understood reasons
the core of HeI 5876 does give consistent r.v. results for our epochs, providing 
the additional decimal place in the period that is needed to phase our data over one year. 

\subsection{Quantitative Characterization of the Wind Evolution}

We tried fitting multiple components to line profiles, but the profiles are
so complicated that our results were quite dependent on the number of 
components chosen.  The fact that blueshifted H$\alpha$ emission
sometimes overlaps blueshifted H$\alpha$ absorption further complicated the
fitting.  Therefore we chose instead the use of direct summations to
extract quantitative measures of the strengths and velocities from the line
profiles, which does not require assuming anything about the profile
shapes or the number of components.  We defined a window between between -750 
and -2500 km s$^{-1}$ (in the 
center-of-mass rest frame) over which to perform numerical integrations.  The
blue edge of the window was chosen so as to capture the bluest portions of the
profile features due to the wind, while the red edge was fixed so as to try to 
avoid the strong central emission.  (Because the central emission line and the
wind features sometimes overlap, the separation is not perfect, but is largely
effective).  All pixel
values above the continuum were assumed to belong the emission component, and those
below to absorption.  These integrations were converted to equivalent widths (EW),
which were taken as  measures of the emission and absorption strengths, and the first 
moments of the distributions were taken as the characteristic wavelength (i.e., velocity) 
of the absorption and emission features.  That is,

$EW_{em} = \Sigma(1 - I_{i})\Delta\lambda$, summed over the window, for $I_{i}>1$,

$EW_{abs} = \Sigma(1 - I_{i})\Delta\lambda$, summed over the window, for $I_{i}<1$,

$\lambda_{em}  = \Sigma \lambda_{i}(I_{i} - 1)/\Sigma \lambda_{i}$,
summed over the window, for $I_{i}>1$, and

$\lambda_{abs} = \Sigma \lambda_{i}(1 - I_{i})/\Sigma \lambda_{i}$, 
summed over the windown, for $I_{i}<1$.

Note that both absorption and emission are
measured from the same wavelength window.  Also note that because the continuum will
have some noise, there will always be some pixels above the continuum, leading
to a small but finite emission line EW even in the absence of an emission feature,
with a corresponding effect in absorption.

Figure 6 shows the results of these measures for each of the nine sequences,
where we have plotted vs. time the r.v. and EW of the blueshifted absorption
component of both HeI 5876 and H$\alpha$ (top two panels).  The bottom panel
plots the EW vs time for the blueshifted emission of H$\alpha$.  We have not plotted
the velocities of the blueshifted emission of H$\alpha$ because the feature is 
usually very broad with mostly random mean velocity changes.  Also, velocities are
not plotted when the line is weak and near the detection limit.  The errors on the EWs 
and velocities can be judged by the scatter in the plots during times when the
feature is not varying or (in the case of the EWs) when the feature is at or near
zero.  EW errors are typically 0.1-0.3\AA\ and the r.v. errors are typically
100-300 km s$^{-1}$

Examining the middle panels of Figure 6 we see that, in general, the velocities of the 
wind absorption decrease with time, approximately linearily.  For 10 relatively isolated
blueshifted absorption events in the top panels of Figure 6, which are also strong
enough to have radial velocity measurements in the middle panels of Figure 6, we
find 3 events with weak or ill-defined velocity evolution, 1 event in which the absorption
becomes bluer, and 6 events for which the absorption evolves towards lower expansion
velocities.  RN98 also concluded that most BZ Cam wind events decelerated with time.  This 
trait is therefore a frequent and persistent property of the wind episodes in BZ Cam,
but contrary behavior is sometimes seen.
 
\subsection{Orbital Dependence of the Wind}
 
The EW of the blueshifted HeI 5876 absorption appears to be the most 
sensitive and reliable indicator of wind activity in BZ Cam.  It is generally
stronger than the blueshifted H$\alpha$ aborption and, unlike H$\alpha$,
is not contaminated by overlapping emission from the wind.  Using the
HeI EWs shown in the top panels of Figure 6 we have plotted these measures
of wind strength vs. orbital phase in Figure 7.  Data points belonging to
same spectral sequence are connected by straight lines, showing that the
episodic nature of the wind occurs at all orbital phases.  Nevertheless
the episodes are clearly concentrated to phases near or just following
superior conjunction of the mass-losing star.     

\subsection{Simultaneous Photometry and Spectroscopy}

On three nights we were able to obtain simultaneous V band photometry to
accompany the spectroscopy.  Figure 8 compares the V-band brightness of the system
to the strength of the wind, as measured by the EW of blueshifted absorption in
HI 5876.  In the top panel it appears that all of the undulations are in common
to the continuum brightness and to the strength of the wind.  (It is unfortunate 
that the strong wind event at the end was missed in the photometry.)  The middle 
panel shows no correlation between the
photometry and the wind.  In the bottom panel the slow rise over the middle portion
of the observing window is in common to the photometry and to the wind strength, but
there are no photometric features corresponding to the wind enhancements at the
beginning and end of the sequence.

\section{Discussion}

RN98 discuss how the apparent deceleration during a BZ Cam wind event is 
(surprisingly) opposite to the acceleration seen in the winds in OB stars, whose
line profiles resemble those of NL CVs.  They suggest that the effect may be due
to wind dilution as it expands.  We think that the apparent slowing of the
wind velocities in BZ Cam is a geometrical effect due to the component of motion 
of the absorbing wind across the accretion disk as a wind event progresses.   Imagine 
that a wind event involves the ejection of one (or a few) blobs of gas having a 
range of speeds, at a moderate opening angle.  This will produce a range of
transverse velocities of the absorbing wind blobs, such that those 
wind components with
larger transverse velocity will complete their motion across the face of the disk
before those components with smaller transverse velocity.  Those faster components
will also produce greater blueshifted absorption as long as the wind remains projected
in front of the continuum source of the disk.  Our evidence for this scenario is
that the line profile changes seen during the velocity decline are such that the
high velocity absorption disappears first, leaving only low velocity absorption
late in the event.  That is, the profile change is not one in which a given
range of absorption velocities shift together to the red, but rather one in
which the selective removal of the most blueshifted absorption as the event progresses
leads to a net redward shift.  At our time resolution the wind events seem to
almost always begin with a very broad shallow absorption, which evolves to  profiles
having deeper, more narrow, and less blueshifted absorption.  This systematic 
change in line profile can be seen in most of the wind episodes displayed in Figure 2,
when the S/N is favorable.  For example, note the He I 5876 line profile changes 
in spectra 1-01 to 1-08, 3-01 to 3-06, 3-50 to 3-59, 4-09 to 4-16, and 5-09 to
5-27.  Typically the velocity has declined in 
$\sim$40 min before disappearing due to weakness and/or merger with the central
emission line.  Following RN98, a disk radius of 4$\times$10$^{10}$ cm (characteristic of
luminous CVs) will be traversed in 40 min for a transverse velocity of 170 km s$^{-1}$.
Considering the uncertainties in the wind opening angle and the fraction of the
disk traversed during a wind event, this rough agreement seems reasonable.

In the scenario just outlined, the blueshifted emission components of H$\alpha$ seen in 
some of our spectra might result from gas that has passed beyond the edge of the disk 
continuum source, thereby changing from absorption to emission.    
If this interpretation of the
nature of the blueshifted H$\alpha$ emission components is correct, then 
we would expect to find that blueshifted H$\alpha$ emission components systematically
appear after a P-Cygni absorbing episode in HeI 5876.
Unfortunately our spectral sequences are not long enough or extensive enough for a
definitive test, but we can nevertheless make a quantitative consistency check.  
There is an interval following the start of a HeI 5876 P~Cygni absorption episode during
which we have the opportunity to check for blueshifted H$\alpha$ emission that might
result from the late stages of the wind event.  This window of opportunity (WOO) may
be terminated by either the beginning of new wind episode, or by the termination of
the spectral sequence.  The typical spacing between wind episodes is only $\sim$80 min,
which provides only a few useable WOO's having well-defined pairs of HeI blueshifted 
absorption events and blushifted H$\alpha$ emission events.  In Figure 6b (spectral 
sequence S2) we find a delay of 29 min between
the start of the HeI 5876 absorption and the start of the blueshifted H$\alpha$ emission,
within a WOO of 172 min.  Therefore the blueshifted H$\alpha$ emission occurs 
after a fraction 0.16 of the TOO has elapsed.  There are two such pairings in Figure 6e
(spectral sequence S5), yielding ratios Delay/WOO of 36/115 = 0.31 and  43/94 = 0.46. 
Because our sample is so small, a probability discussion is not appropriate.  Instead
we simply note that, compared to a random distribution of blushifted H$\alpha$ emission 
within a TOO, all 3 occurances are early in the TOO, consistent with our hypothesis. 
Furthermore, the measured delays are consistent with estimates of disk transit times. 

Similarly, the {\em redshifted} H$\alpha$ emission components in the RN98 spectra might be 
due to wind concentrations on the {\em backside} of the disk, after they have passed out 
from behind the disk.  In that case the passage of a backside wind blob across the
disk does not produce a P~Cygni aborption event, so we expect no correlation of
HeI 5876 absorption with redshifted H$\alpha$ emission.  We attempted an analysis of
the RN98 results in a fashion similar to that just described for the blueshifted H$\alpha$
components in our spectra.  However, difficulties with small numbers of useable
pairings rendered this attempt futile.  We do note that RN98 concluded that the
red emission wing of H$\alpha$ was uncorrelated with the P~Cygni events in their data,
which is consistent with our suggested scenario.  

There have been numerous tests for an orbital dependence of the wind in NL CVs using
space UV lines, with mixed results.  This work has been hampered by various combinations of
limited S/N, limited time resolution, and limited number of orbits.  The limited number of
orbits is a particular handicap for those systems in which the finite length of wind episodes
can mimic a dependence on orbital phase.  Taken together these studies (listed in Prinja 
et al. 2004
and in Froning 2005) demonstrate that orbit-modulated winds do occur, in spite of the
difficulty in interpretation.  The FUSE observations of the wind in V592 Cas (Prinja et al. 2004) 
is a study in which the orbital modulation is very secure.  Furthermore this study was
able to rule a wind modulation on either the positive or negative superhump period in 
V592 Cas, implying that neither disk eccentricity nor tilt is responsible for the orbital
modulation.  The orbital modulation of the wind in RW Sex (Prinja et al. 2003) provides further 
evidence that phase-dependent variations in CV winds are not tied to disk tilt or to disk 
eccentricity, because SW Sex is not a superhump system.
Insofar as tests for an orbital dependence of the wind in BZ Cam itself, the IUE 
study of Griffith, Fabian \& Sion (1995) provided only a suggestion of such an effect, while the
HST UV spectral study of Prinja et al. (2000) did not find any conclusive evidence for
phase modulated absorption changes in the wind.  Optical tests for the phase dependence
of the winds in NL CVs are rare, but Kafka et al. (2009) found that the wind events
in V592 Cas seemed to be concentrated near phase zero (- to + crossing
of $\gamma$ for the HeI emission lines, or superior conjuction of the secondary star).  
This is the same phase range in which we find the BZ Cam wind episodes to be concentrated.  

Prinja et al. (2004) discuss the puzzle of an outflow modulated on the orbital
period, reviewing the observations and suggested solutions.  After ruling out eccentric or 
tilted disks, one of the few mechanisms remaining 
as possibilities for breaking the axisymmetry of the wind in V592 Cas (and by implication 
in the winds of NL CVs as well) is effects arising where the accretion stream meets the 
disk at a particular
disk azimuth.  Stream impact velocities are $\sim$5-10$\times$ smaller than
wind velocities, so it is unlikely that the stream/disk interaction can produce a wind 
directly.  Instead, we propose that stream impact and penetration raises gas clumps far 
enough above the plane of the disk that radiation pressure from inner disk can accelerate 
these clumps to the observed outflow velocities.  In this scenario the stream/disk impact 
serves to imprint the observed episode time scales onto the wind (from the larger amplitude 
portion of the flickering), and also produces an orbital modulation.

Time dependent models of radiation driven winds from luminous accretion
disks (Proga, Stone \& Drew 1998) produce stochastic velocity and density fluctuations 
in the wind with characteristic time scales of a few hundred minutes, if the
radiation field arises mostly from the disk.  
The cause is the difference in the variation with height of the vertical components of gravity 
and radiation force.  The fluctuating, clumpy, turbulant layer is concentrated near the 
plane of the disk,
at low expansion velocities, whereas the BZ Cam wind variations encompass a wide range
of velocities.  However, the time scales of the variations produced by these models are 
intrigingly similar to the wind episode time scales seen in BZ Cam.  The wind 
variablility mechanisms
described in Proga et al (1998) may well be in play for CV winds like those in BZ Cam, even if
other mechanisms (such as the hotspot-assisted radiation driven wind we are proposing)
leads to the the observational effects discussed in this paper.  Note that the Proga et al. 
models have cylindrical symmetry and therefore cannot address the orbital modulation
of the wind, which is integral to our suggested scenario.     

In Figure 7 we see that the strongest blueshifted wind absorption occurs near orbital 
phase zero, or superior conjunction of the secondary star. 
If the stream/disk interaction is responsible for launching the localized
density enhancements that become the wind episodes, then does this orbital phase make sense?  
In high inclination CVs a 
"hump" in continuum light often appears just before inferior conjunction of the secondary
star, arising from the "face-on" presentation of the impact point of the accretion stream 
with the accretion disk.  (Actually, while hot spot humps are common in high inclination 
dwarf nova CVs  they are rare in NL CVs such as BZ Cam, a complication we can ignore 
for now.)   In high inclination NL CVs there is often evidence of stream/disk 
interactions on the far side 
of the disk, which can happen when the stream is thicker than the disk and overflows the initial 
impact point.  This overflow can re-impact the disk on the far side, producing an
interaction site that is at a quite different disk azimuth than the original disk impact
azimuth, and often much nearer to the inner disk.  These effects have been explored theoretically 
in studies such as Lubow (1989)
and Armitage \& Livio (1996; 1998).  They have also been studied observationally, mostly
in the context of the SW Sex phenomenon (e.g. Szkody \& Pich\'{e} 1990; 
Hellier \& Robinson 1994; Knigge et al. 1994; Hoard et al. 1998).  In general,
these studies have found that the second stream/disk interaction can take place at a variety
of disk azimuths and radii which are well inside the outer edge of the disk and mostly on the
far side of the disk from the initial stream impact point.  Therefore we do not think that
the phase of maximum wind strength in Figure 27 is particularly significant, especially 
since in this low inclination system the structures are mostly visible at all orbital phases.
Models of stream/disk interaction sites by Armitage \& Livio (1996) can produce a
"spray" of debris above the disk.  It was speculated that this gas may break up into clouds 
under the influence of radiation from
the central source (or clumps may be imposed by irregularities in the stream).  
We speculate that such clumps are then accelerated by radiation pressure, imposing 
the flickering time scales on the wind episodes as well introducing departures from 
axisymmetry in the wind.  In BZ Cam the flickering power extends
to frequencies as low as the typical wind episode intervals (Patterson et al. 1996).
Therefore a comparison of the V magnitude to the wind episodes bears on the feasibilty
of the scenario just proposed. 

Figure 8 shows that system brightness (V magnitude) and wind strength
(as measured by the EW of blueshifted absorption in HeI 5876) are sometimes, but not always,
correlated with one another.  The data in the top panel shows good time correlation between 
the two measures.  In the region of overlap nearly all of the inflections are in
common to both the wind and the flickering brightness.  In the middle panel there is
no correlation between wind strength and flickering, while in the bottom panel there
is some agreement over the midrange, but the two wind episodes have no counterpart in
the photometry.  Overall it is seen that the timescales for the wind changes and the
flickering are quite similar, even if not always well correlated.  It is easy to imagine that
complexities in the details of the stream/disk interaction could result in some flickering 
events primarily producing only backside wind events, leaving no P~Cygni signature.
This speculation might account for the disappointing degree of correlation in the bottom 
two panels of Figure 28.

The process we have outlined (a hot spot assist to a radiation-driven wind) adds an additional
mechanism to the usual scenario for wind generation.   Nevertheless we think that
a confrontation with Occam's razor is not unfavorable, because the proposed scenario 
addresses two additional observational properties: 1) the episodic nature of the winds
in NL CVs,  and 2) the orbital dependence of the wind.  This proposed scheme is schematic 
and qualitative at this point, and will need additional observations and modeling (which
are outside the scope of this paper) to establish its true merit.  

\section{Summary and Conclusions}

Our primary observational results and suggested interpretations can be summerized 
as follows:

1) The wind episodes in BZ Cam typically last 35 min, during which a systematic
   decline of the velocities of the blushifted absorption occurs.  This
   decline is attributed to the motion of wind concentrations across the face
   of the disk, whereby the blobs producing the largest blueshifts are more rapidly 
   lost from the absorption line profile as they move beyond the source of
   background continuum light from the disk.
   
2) We find numerous examples of high velocity blueshifted emission components
   to the H$\alpha$ line profile, which we attribute to emission from blobs
   of wind on the frontside of the disk after the gas concentrations have passed
   the outer edge of the disk.  These features are in contrast to similar high 
   velocity {\em redshifted} emission components in the H$\alpha$ line profile 
   found in RN98, which may be due to analogous backside wind events.  We
   fnd that the delays between the start of a wind episode and the appearance
   of a H$\alpha$ emission component is consistent with this scenario, but
   this result is not definitive because of small sample sizes.
   
3) We provide a new orbital ephemeris for BZ Cam, using the central core of the
   HI 5876 line.  This ephemeris has sufficient accuracy to correct our line
   profiles for orbital motion and to reveal that the BZ Cam wind events are
   concentrated near the time of superior conjunction of the secondary star.
   We find that the orbital variations of the emission line radial velocities
   are inconsistent from line-to-line, and only the core of HeI 5876 gives
   good results.
   
4) Simultaneous photometric and spectroscopic sequences on three nights found
   that the flickering was well correlated with the wind on one night, somewhat
   correlated on a second, and uncorrelated on a third night.  The timescales
   for the flickering and for the wind variability were similar.
   
5) Because the flickering is (at least sometimes) correlated with
   the wind, and the wind events are concentrated at a particular orbital
   phase, we suggest that the BZ Cam wind is generated by radiation pressure
   acting on gas blobs that are initially raised above the disk by the
   interaction of the accretion stream with the disk.  In this way the stream/disk
   interaction can imprint upon the wind both the observed flickering time scales
   and the observed orbital dependence of the wind episodes.

\section{Acknowledgments}

We are happy to acknowledge useful conversations with Daniel Proga regarding models
of CV winds, and particularly of his time-dependent wind models.

\begin{deluxetable}{cllccrcc}
\tablenum{1}
\tablewidth{0pt}
\tablecolumns{8}
\tablecaption{Photometry Log}
\tablehead{\colhead{Sequence} & \colhead{UT} & \colhead{JD} & \colhead{Tel/Obs} 
& \colhead{Secs} & \colhead{\# Exps} &\colhead{Dur.} & \colhead{Simul. Sp.}}
\startdata
P1  &  1990-May-16 to & 2448027 to  & IU 0.41-m          & 120	 & 489 & 6.4 yr  &     \\
    &  1996-Oct-14    & 2450370     &                    &	 &     &	 &     \\
    &                 &             &                    &	 &     &	 &     \\      
P2  &  2000-Aug-15 to & 2451771 to  & IU 0.41-m          & 120	 & 508 & 4.5 yr  &     \\
    &  2005-Feb-21    & 2453432     &                    &	 &     &	 &     \\
    &                 &             &                    &	 &     &	 &     \\      
P3  &  2007-Sep-01 to & 2454344 to  & IU 1.25-m          &  90	 & 269 & 4.6 yr  &     \\
    &  2012-Apr-09    & 2456036     &                    &	 &     &	 &     \\
    &                 &             &                    &	 &     &	 &     \\
P4  &  2005-Oct-12    & 2453655     & Tenagra 0.76-m     &  90	 &     & 3.5 hr  & S3 \\      
P5  &  2006-Feb-21    & 2453787     & WIYN 0.91-m        &  60	 & 126 & 5.4 hr  & S7 \\
P6  &  2006-Feb-23    & 2453789     & WIYN 0.91-m        &  60	 & 115 & 4.6 hr  & S8 \\      
\enddata
\end{deluxetable}
\clearpage

\begin{deluxetable}{cccc}
\tablenum{2}
\tablewidth{0pt}
\tablecolumns{4}
\tablecaption{Magnitudes for BZ Cam}
\tablehead{\colhead{JD}  & \colhead{V Mag} & \colhead{Error} & \colhead{Source}}
\startdata
 2448207.73917  &  12.453  &  0.022  &  IU 0.41-m  \\
 2448207.95582  &  12.509  &  0.012  &  IU 0.41-m  \\
 2448208.82453  &  12.471  &  0.011  &  IU 0.41-m  \\
 2448209.68741  &  12.386  &  0.019  &  IU 0.41-m  \\
 2448233.74992  &  12.614  &  0.019  &  IU 0.41-m  \\
 2448234.63977  &  12.445  &  0.013  &  IU 0.41-m  \\
 2448234.81600  &  12.629  &  0.024  &  IU 0.41-m  \\
 2448235.61472  &  12.471  &  0.015  &  IU 0.41-m  \\
 2448235.96588  &  12.492  &  0.010  &  IU 0.41-m  \\
.............   & ......   & .....   & .....       \\
.............   & ......   & .....   & .....       \\
.............   & ......   & .....   & .....       \\                                                     
 2456027.59723  &  13.050  &  0.004  &  IU 1.25-m  \\
 2456029.61310  &  12.720  &  0.004  &  IU 1.25-m  \\
 2456036.58051  &  13.042  &  0.003  &  IU 1.25-m  \\
\enddata
\end{deluxetable}
\clearpage

\begin{deluxetable}{ccccccccc}
\tablenum{3}
\tablewidth{0pt}
\tablecolumns{9}
\tablecaption{Spectroscopy Log}
\tablehead{\colhead{Sequence}  & \colhead{UT} & \colhead{JD} & \colhead{Tel}
& \colhead{Secs} & \colhead{\# Exps} & \colhead{Hrs}
& \colhead{Phot?} & \colhead{Clouds} }
\startdata
S1 & 2005-Oct-10 & 2453653 & KPNO 2.1-m & 300     &  8 & 0.8 &     & clear \\
S2 & 2005-Oct-11 & 2453654 & KPNO 2.1-m & 180     & 40 & 2.6 &     & m. clear \\
S3 & 2005-Oct-12 & 2453655 & KPNO 2.1-m & 180     & 59 & 3.9 & P4  & clear \\
S4 & 2005-Oct-25 & 2453668 & WIYN 3.5-m & 300     & 17 & 2.2 &     & clear \\
S5 & 2006-Jan-01 & 2453736 & KPNO 4.0-m & 120-180 & 69 & 3.8 &     & clear \\
S6 & 2006-Jan-03 & 2453738 & KPNO 4.0-m & 180-600 & 22 & 3.3 &     & p. cldy \\
S7 & 2006-Feb-21 & 2453787 & WIYN 3.5-m & 300     & 35 & 4.2 & P5  & p. cldy \\
S8 & 2006-Feb-23 & 2453789 & WIYN 3.5-m & 300     & 26 & 3.5 & P6  & p. cldy \\
S9 & 2006-Sep-15 & 2453993 &  MMT 6.5-m &  90     & 40 & 1.3 &     & clear \\
\enddata
\end{deluxetable}
\clearpage


\begin{figure} 
\figurenum{1} 
\epsscale{0.9}
\plotone{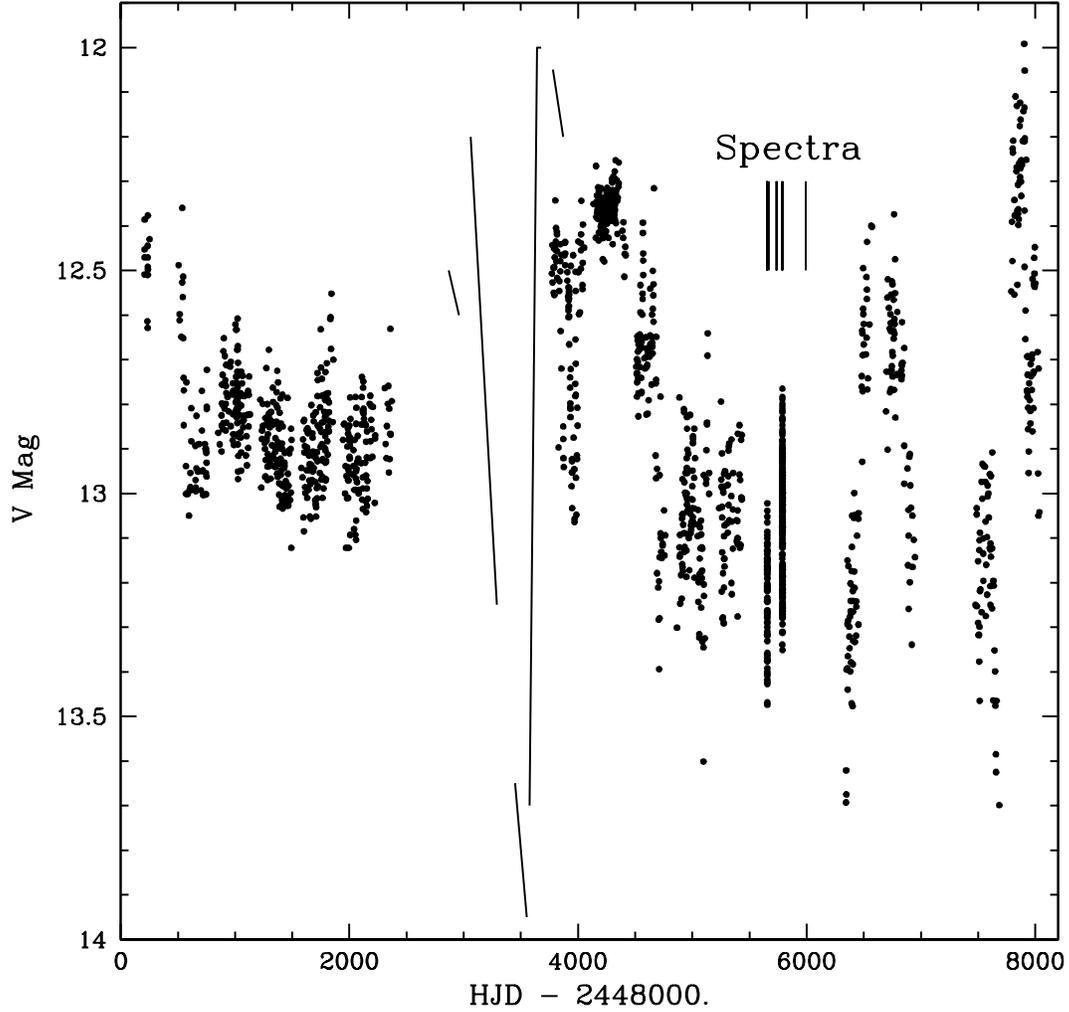}
\caption{The full light curve of BZ Cam from 1990-Nov-12 to 2012-Apr-09.  The vertical 
clusters of unresolved points are continuous sequences of exposures during a night.
In the region 3000-3800 on the JD scale, where we have no original data,
we have added as solid straight lines our characterization of the BZ Cam photometry 
found in Greiner et al. (2001), which were compiled from a variety of sources.  The
times of the 9 spectral sequences are marked with vertical lines.} 
\end{figure}
\clearpage  

\begin{figure}  
\figurenum{2a} 
\epsscale{0.9}
\plotone{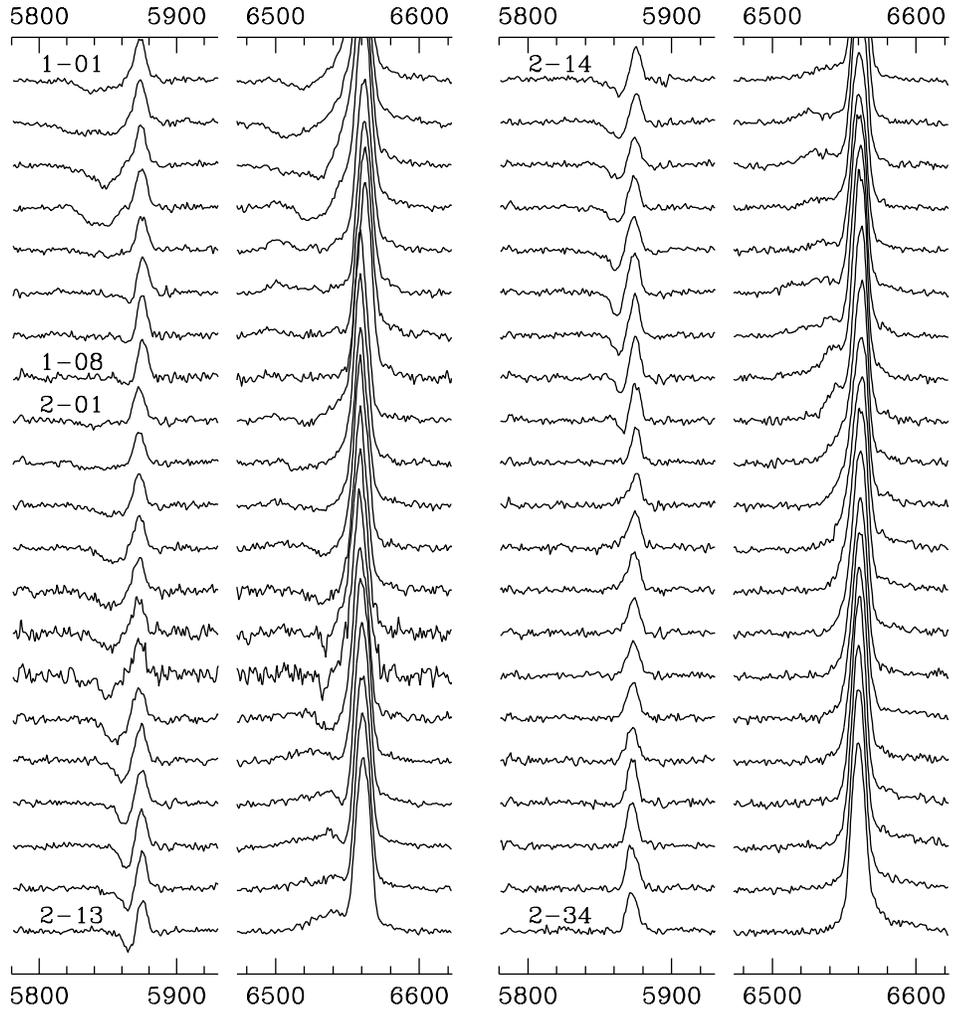}
\caption{Nested plots of the line profiles of HeI 5876 and H$\alpha$.
Time runs down the page.  A profile is labeled if it begins/ends
a column or begin/ends a sequence.}
\end{figure}
\clearpage  

\begin{figure}  
\figurenum{2b} 
\epsscale{0.9}
\plotone{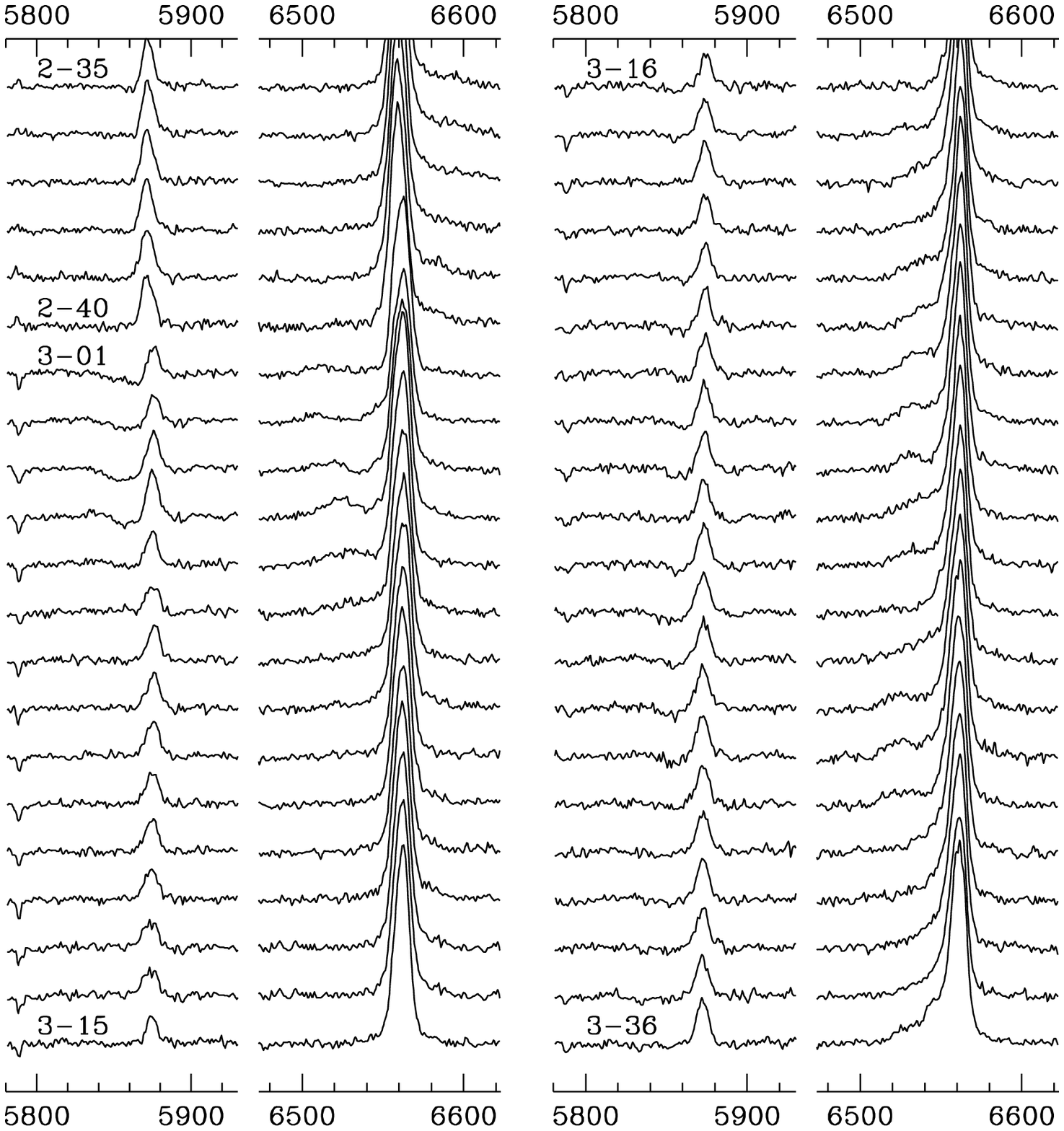}
\caption{Continuation of the line profile plots.}
\end{figure}
\clearpage  

\begin{figure}  
\figurenum{2c} 
\epsscale{0.9}
\plotone{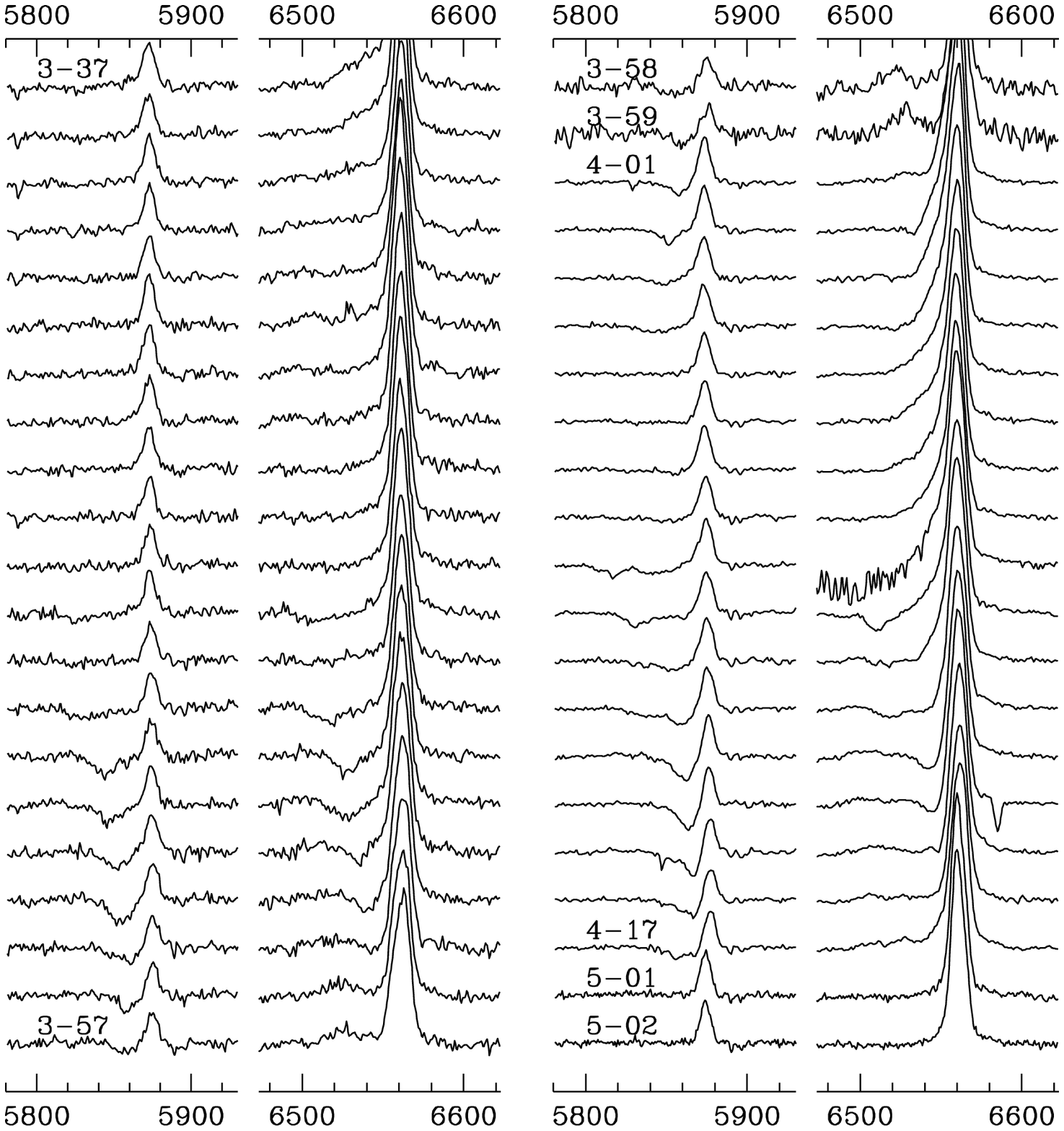}
\caption{Continuation of the line profile plots.}
\end{figure}
\clearpage  

\begin{figure}  
\figurenum{2d} 
\epsscale{0.9}
\plotone{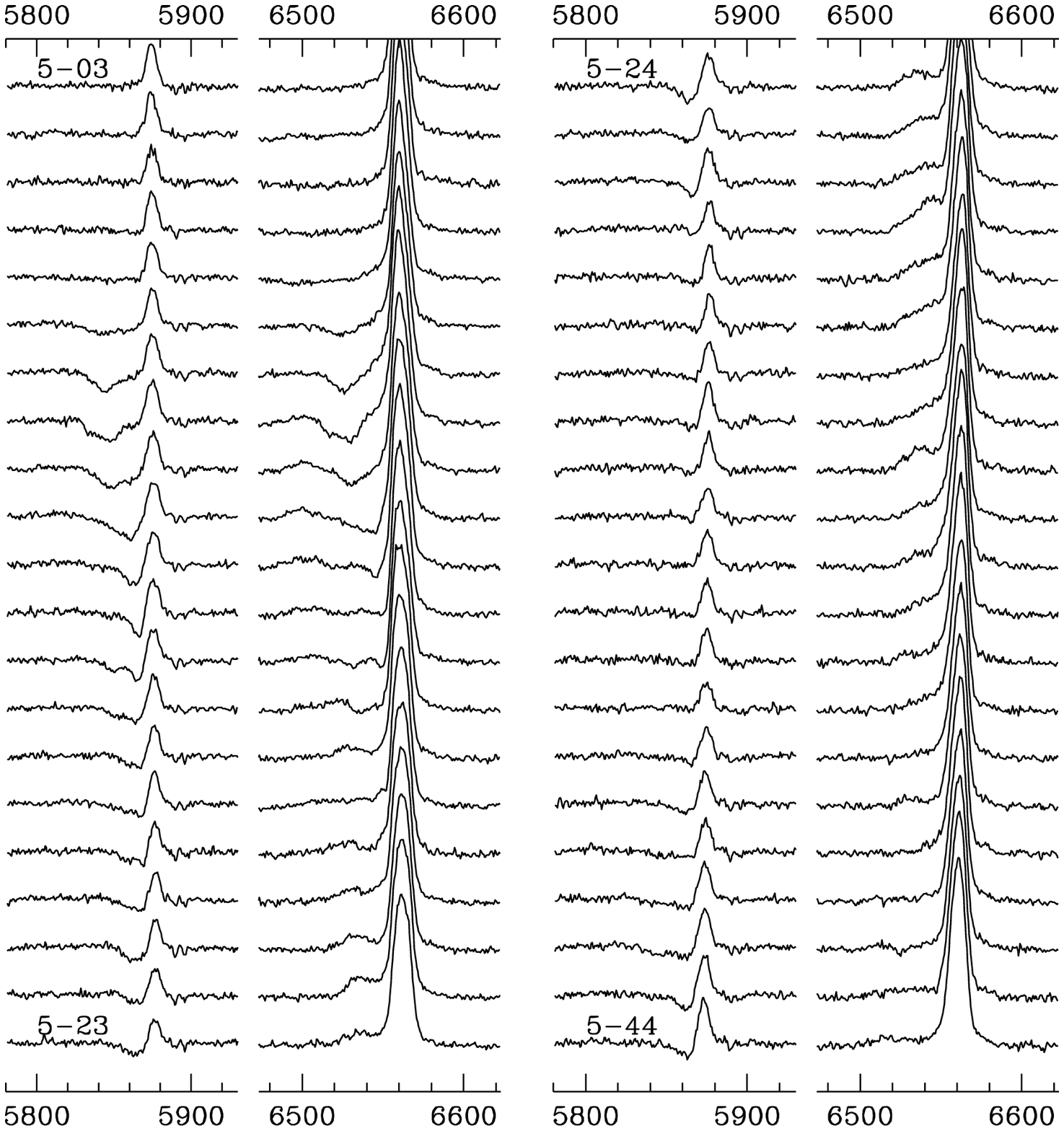}
\caption{Continuation of the line profile plots.}
\end{figure}
\clearpage  

\begin{figure}  
\figurenum{2e} 
\epsscale{0.9}
\plotone{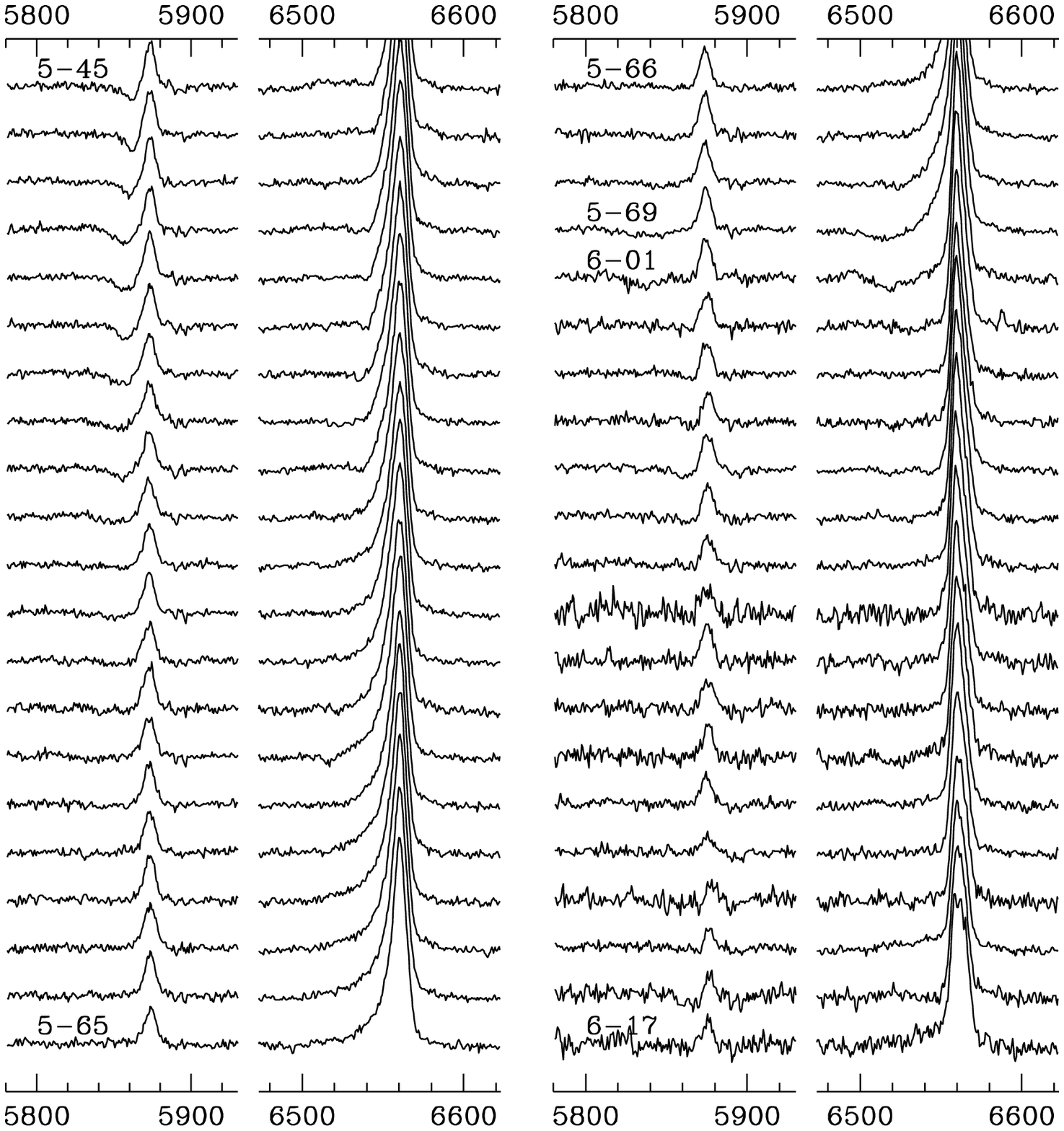}
\caption{Continuation of the line profile plots.}
\end{figure}
\clearpage  

\begin{figure}  
\figurenum{2f} 
\epsscale{0.9}
\plotone{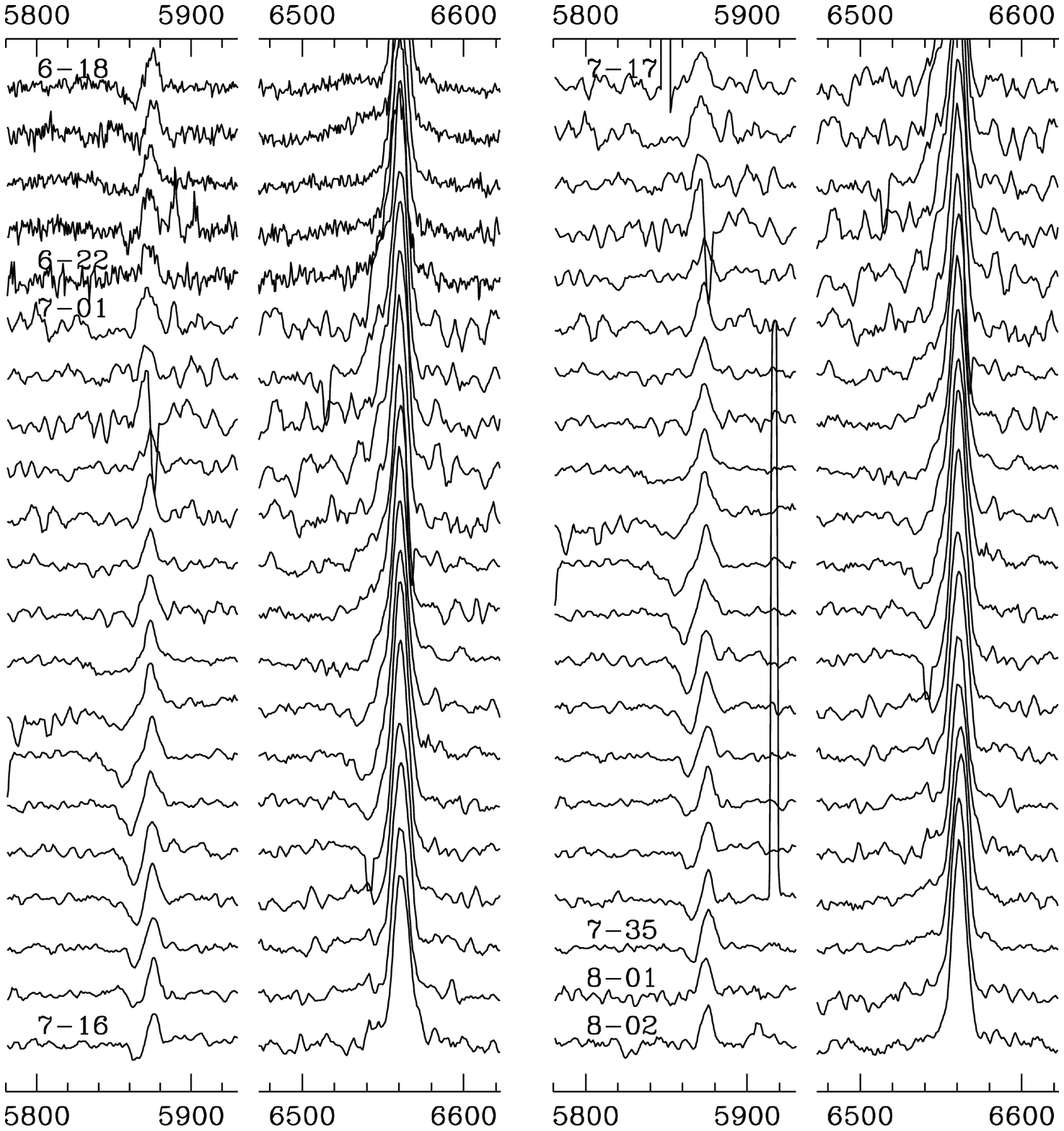}
\caption{Continuation of the line profile plots.}
\end{figure}
\clearpage  

\begin{figure}  
\figurenum{2g} 
\epsscale{0.9}
\plotone{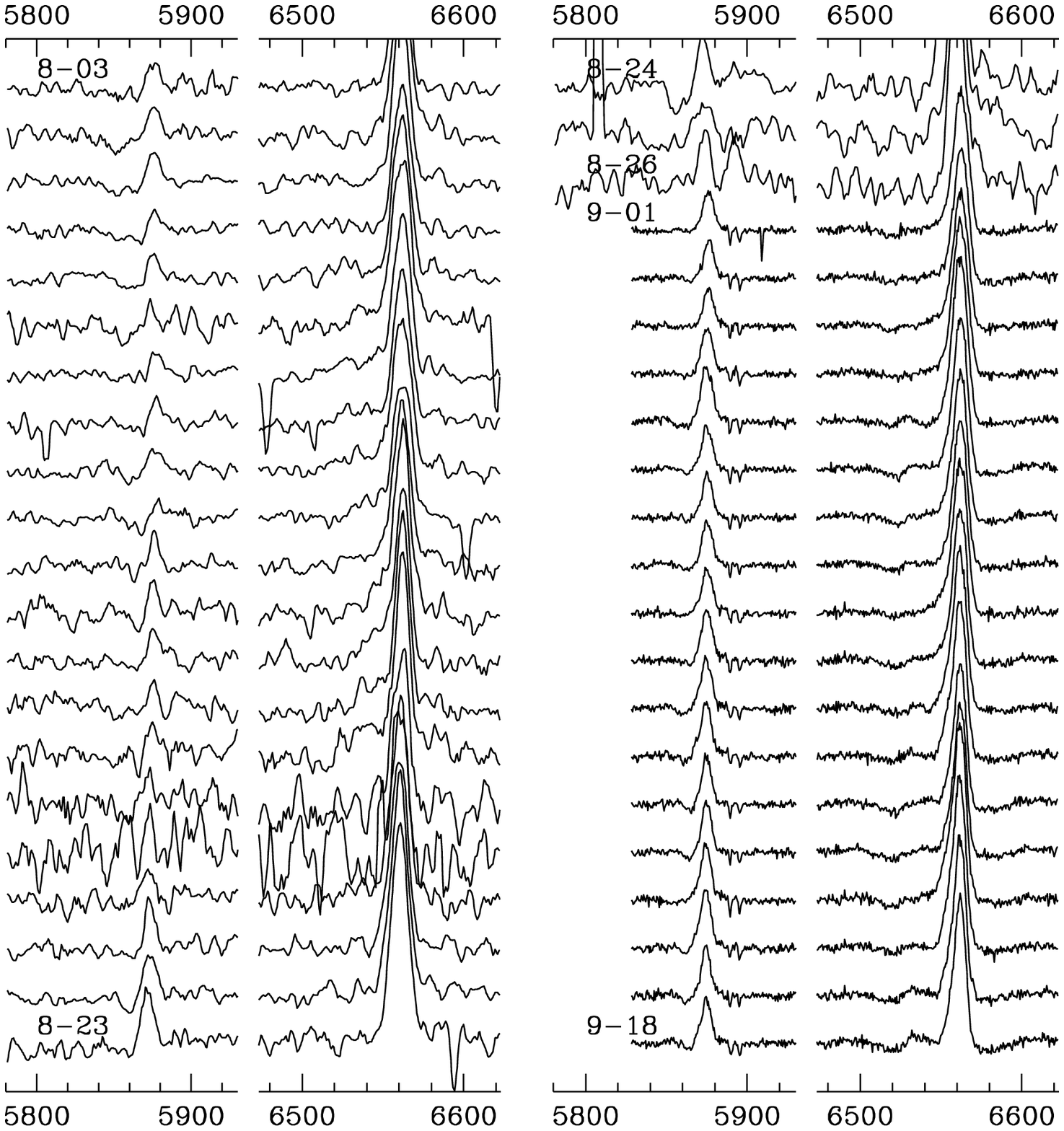}
\caption{Continuation of the line profile plots.}
\end{figure}
\clearpage  

\begin{figure}  
\figurenum{2h} 
\epsscale{0.9}
\plotone{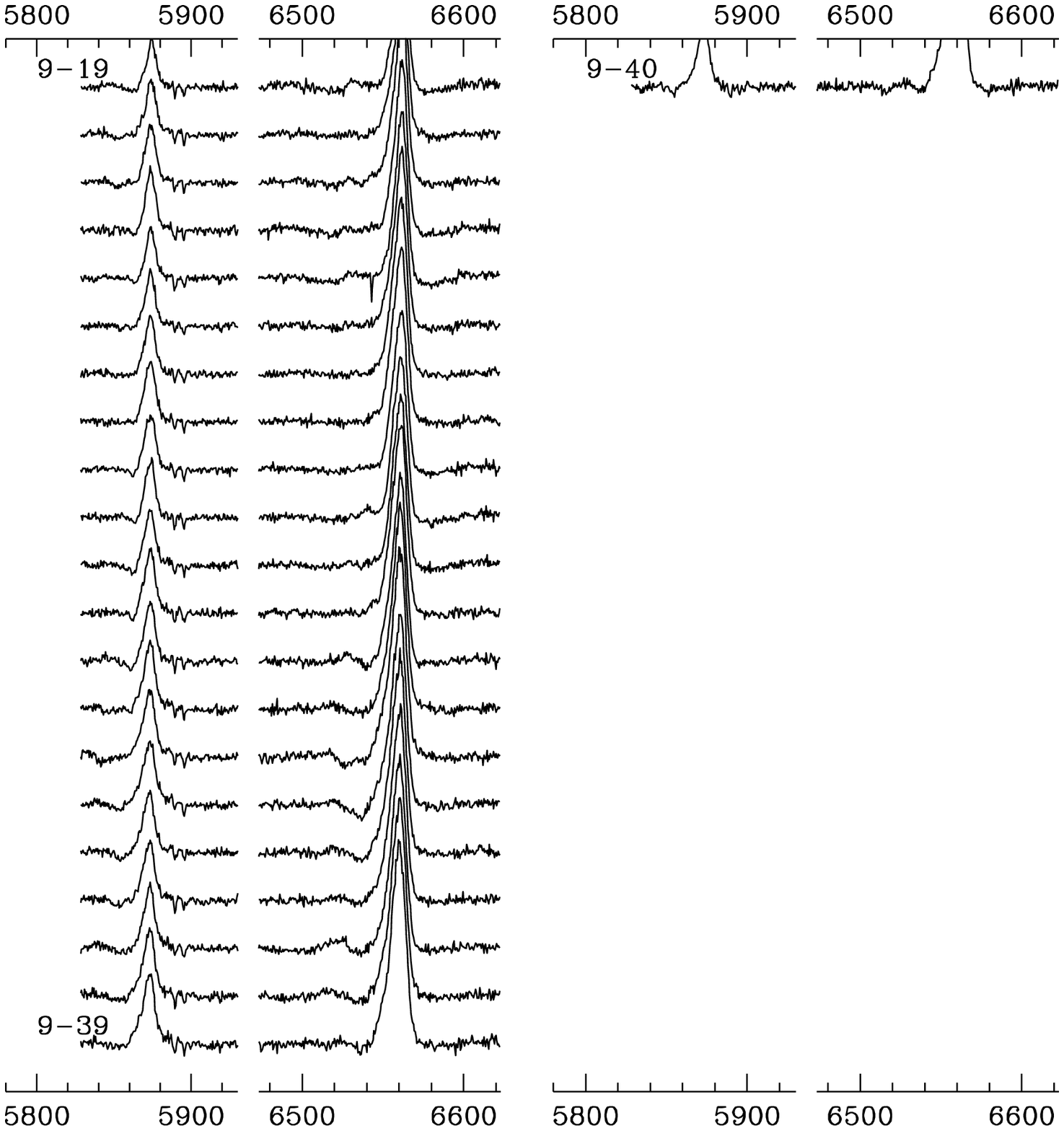}
\caption{Continuation of the line profile plots.}
\end{figure}
\clearpage  

\begin{figure}
\figurenum{3a} 
\epsscale{0.9}
\plotone{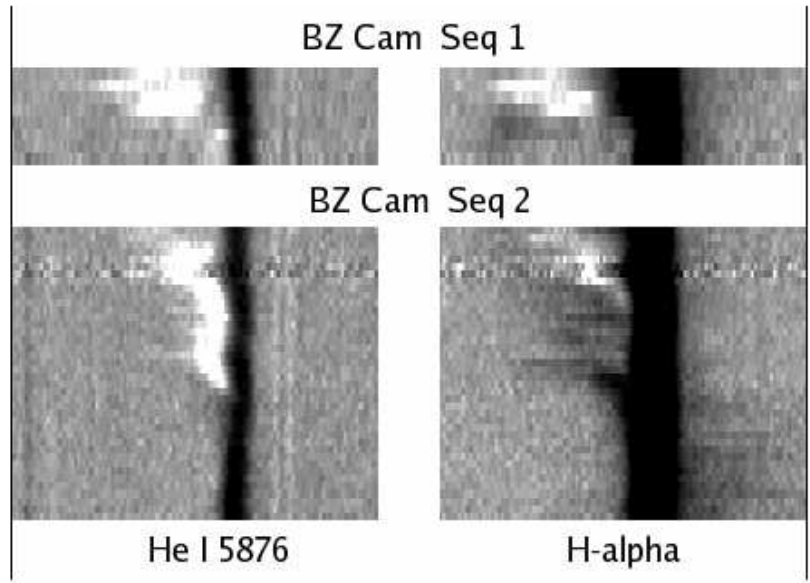}
\caption{Synthetic trailed spectrum for the data of 2005-Oct-10 UT (top) and
2005-Oct-11 UT (bottom).  These spectra are from the KPNO 2.1-m telescope; see Table 3 
and the text for details on these two sequences.  Black is emission, white is
absorption, and time runs down the page.  The intensity scales have been enhanced to 
best show the velocity evolution of wind
features.  Because the exposure times differ from sequence to sequence, the
vertical pixel size has been adjusted so that equal times occupy approximately
equal vertical space on these images.  Because the wavelength per data point differs from
sequence to sequence the data has been interpolated onto a common wavelength scale
covering 150\AA\ around the lines, at 0.5\AA\ spacing.  Some of the sequences were obtained
under varying conditions of cloudiness, which sometimes degrades the S/N of individual
spectra.}
\end{figure}
\clearpage   

\begin{figure}
\figurenum{3b} 
\epsscale{0.9}
\plotone{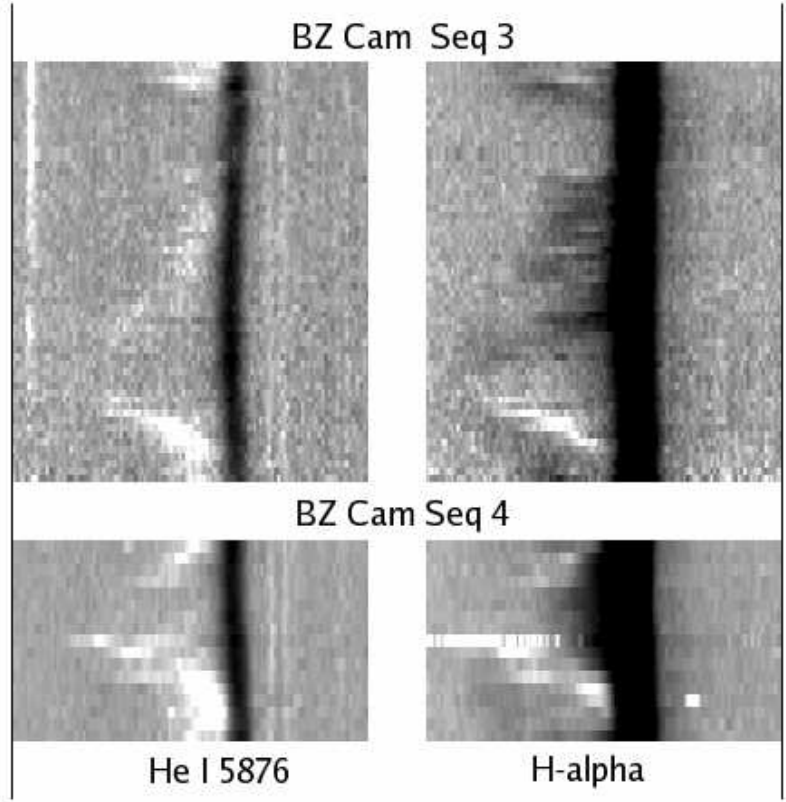}
\caption{Like Figure 3a except for the data of 2005-Oct-12 UT (top) and
2005-Oct-25 UT (bottom).  The top data is from the KPNO 2.1-m telescope, while
the bottom data is from the WIYN 3.5-m telescope.   The apparent strong blue H$\alpha$
absorption structure midway in Sequence 4 is an artifact in spectrum 4-9.   There is
also an artifact just redward of H$\alpha$ in spectrum 4-24.  These features can
easily be seen and identified as artifacts in the nested plots of Sequence 4 
(Figure 5) }
\end{figure}
\clearpage

\begin{figure}
\figurenum{3c} 
\epsscale{0.9}
\plotone{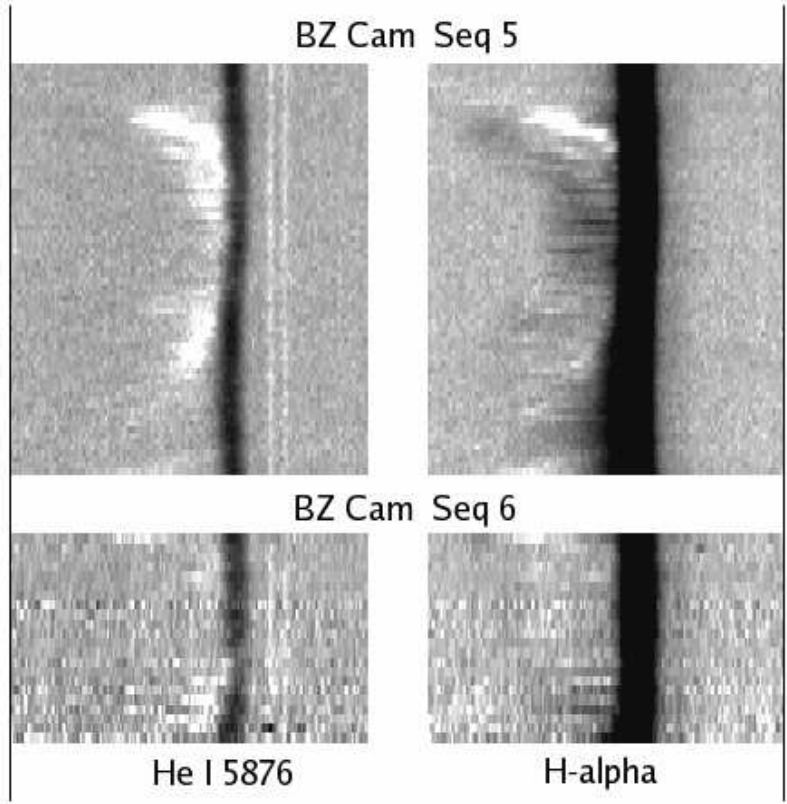}
\caption{Like Figure 3a except for the data of 2006-Jan-01 UT (top) and 2006-Jan-03 UT
(bottom).  These data are from the KPNO 4-m telescope.}
\end{figure}
\clearpage

\begin{figure}
\figurenum{3d} 
\epsscale{0.9}
\plotone{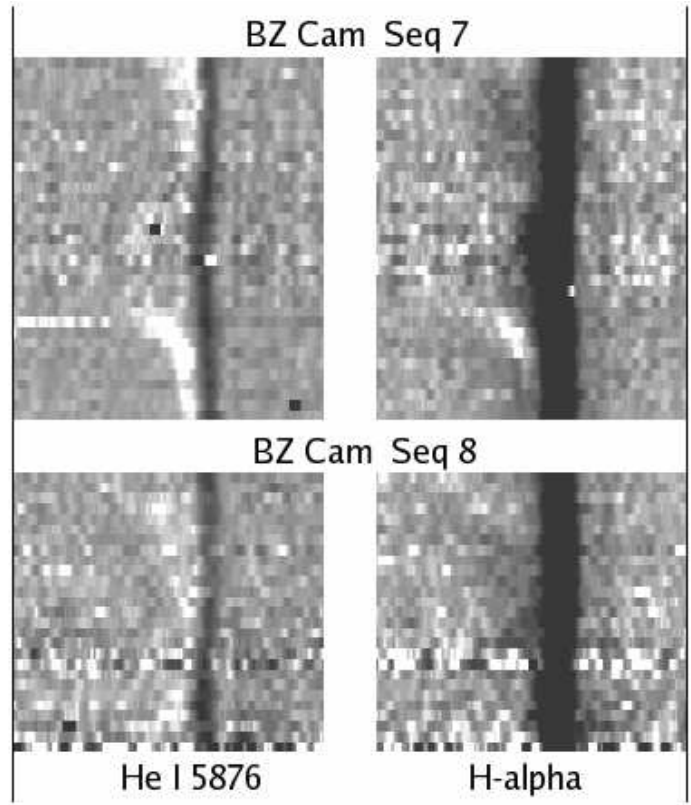}
\caption{Like Figure 3a except for the data of 2006-Feb-21 UT (top) and 2006-Feb-23 UT
(bottom).  These data are from the WIYN 3.5-m telescope.  Because clouds reduced the
S/N of these two sets of WIYN spectra, they have been smoothed with a 3-point triangular
filter.}
\end{figure}
\clearpage

\begin{figure}
\figurenum{3e} 
\epsscale{0.9}
\plotone{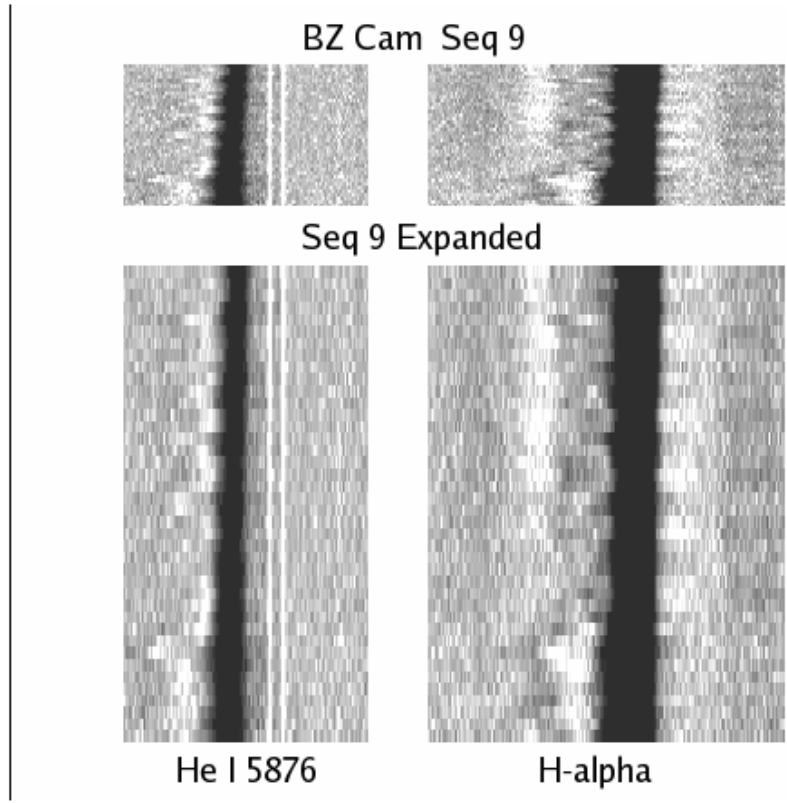}
\caption{Like Figure 3a except for the data of 2006-Sep-15.  These data are from
the 6.5-m MMT.  The top panel has the same vertical time scale as in Figs 11-14.  
Because the exposure times for this MMT sequence are quite short, the bottom panel shows
the same data, but expanded in the time direction by a factor of 3.3.  The region of
spectrum to the blue of the He I line is less than in Figs 11-14 because of the blue
wavelength cutoff of the original MMT spectra.}
\end{figure}
\clearpage

\begin{figure}  
\figurenum{4} 
\epsscale{0.9}
\plotone{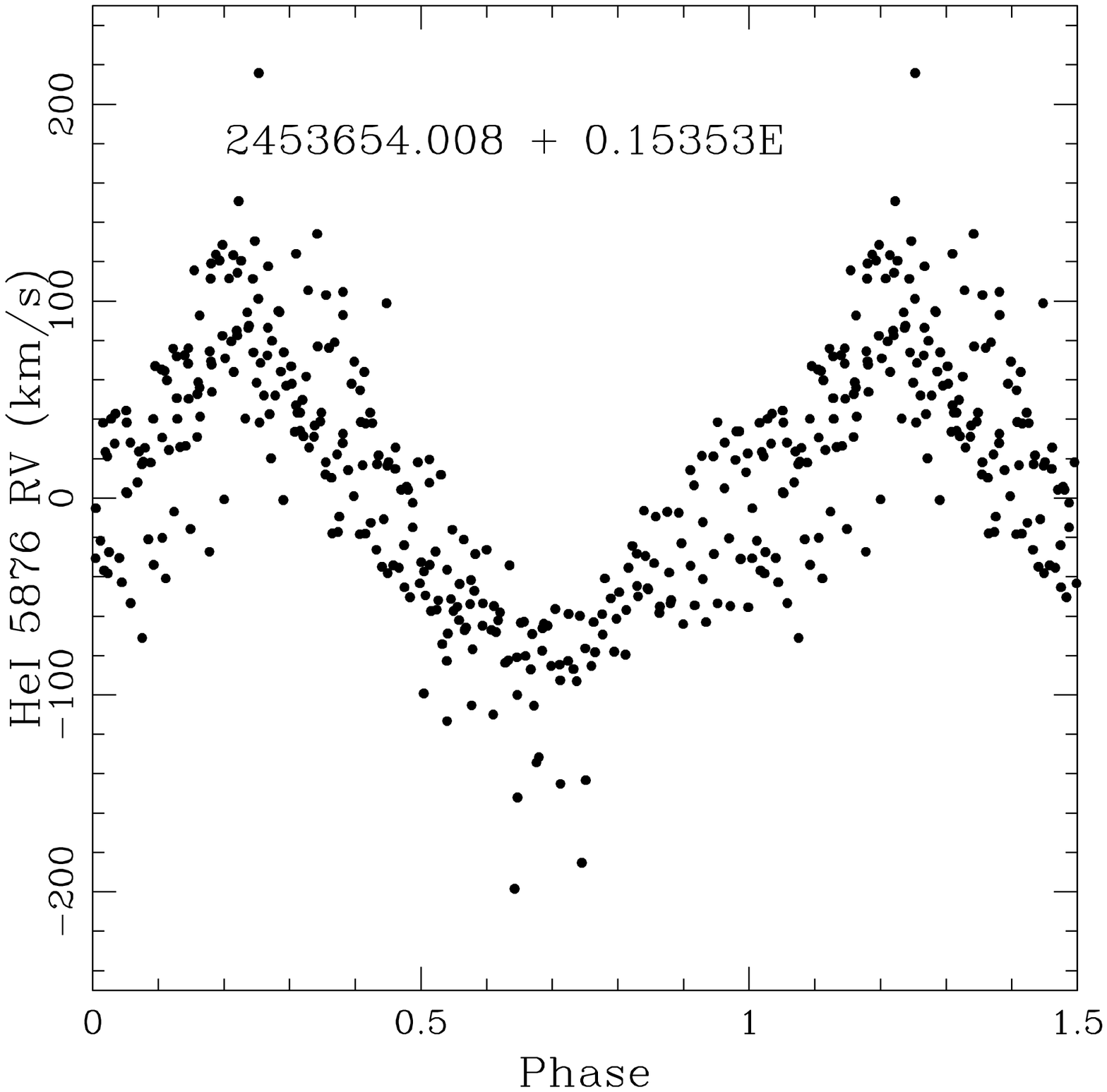}
\caption{The radial velocity curve of the core of HeI 5876, folded on our
adopted orbital period.  Each night's data was pre-whitened prior 
to folding.  The ephemeris is for the time of - to + crossing of gamma.} 
\end{figure}
\clearpage  

\begin{figure}  
\figurenum{5} 
\epsscale{0.9}
\plotone{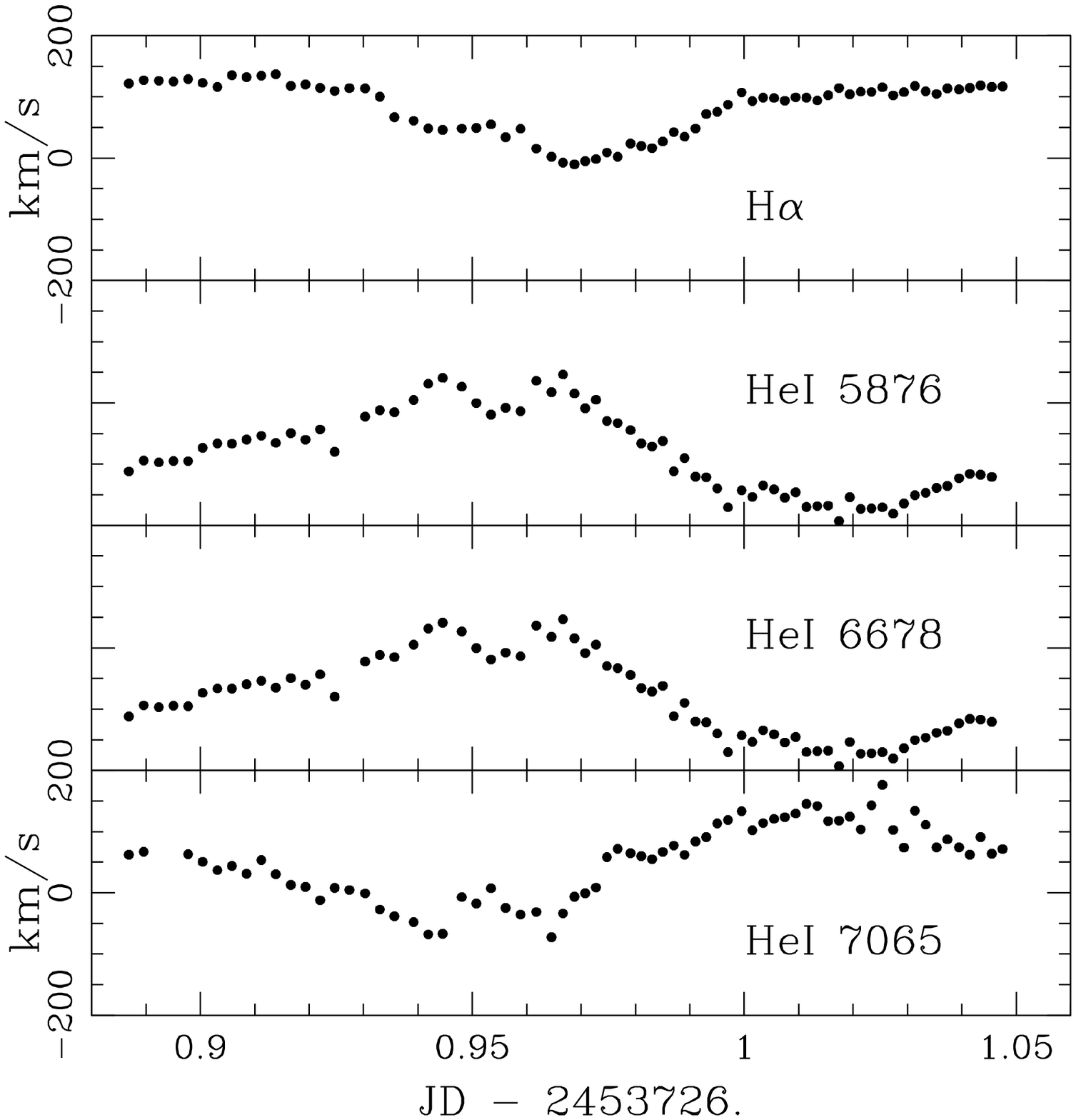}
\caption{The individual radial velocity curves for the cores of 4 emission lines
in BZ~Cam for the night of 2006-Jan-01 (UT) (Sequence S5 in Table 2) showing
the lack of agreement in the shape, amplitude, and phasing of the r.v. curves.
These data have not been pre-whitened.  See text for details.} 
\end{figure}
\clearpage

\begin{figure}  
\figurenum{6a} 
\epsscale{0.9}
\plotone{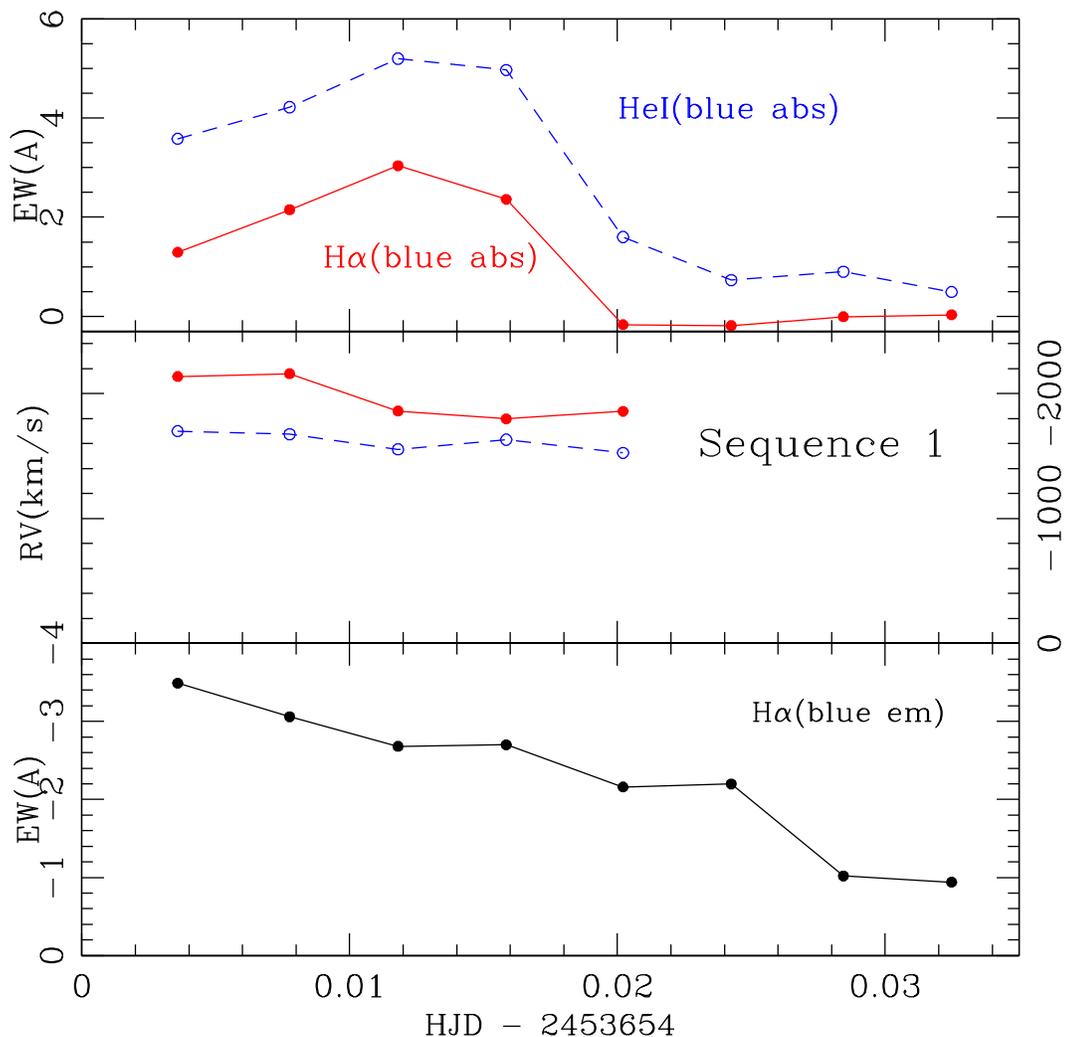}
\caption{Top panel: Wind strength vs time, as measured by the EW of the P-Cygni
absorption in two lines (HeI 5576 (open blue circles)and H$\alpha$ (filled red circles)), 
for spectral set S1 (see Table 3).  Middle panel: Wind velocity as measured by the
radial velocity (km s$^{-1}$) of the blue absorption, using the same symbol coding
as the top panel.  Note that the velocity is missing when the feature becomes too
weak. Bottom panel: The strength of the blueshifted emission component to H$\alpha$
which often accompanies a wind event.   See text for details on how these
quantities are measured.} 
\end{figure}
\clearpage  

\begin{figure}  
\figurenum{6b} 
\epsscale{0.9}
\plotone{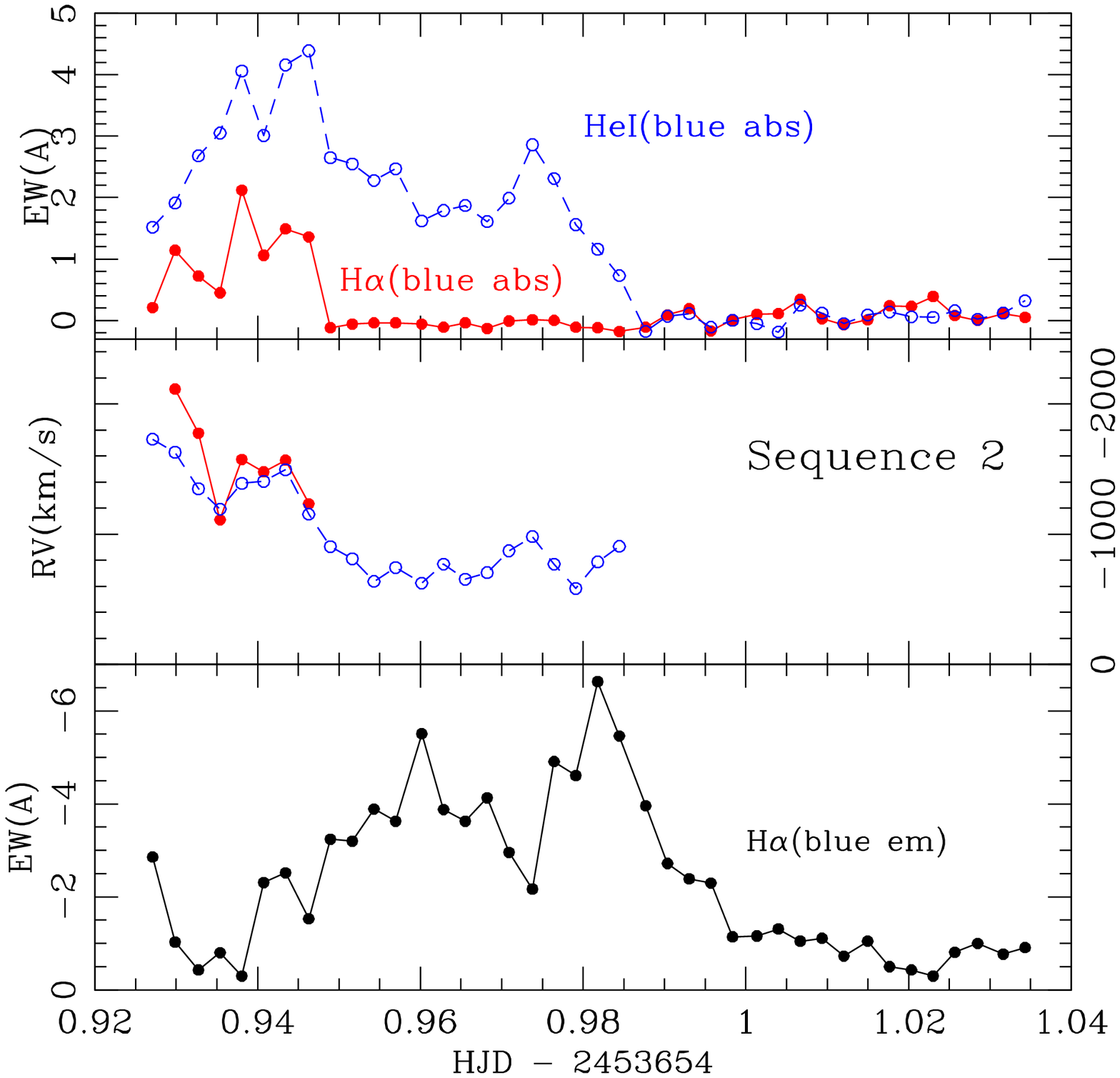}
\caption{Same as Figure 6a, except for spectral set S2.} 
\end{figure}
\clearpage  

\begin{figure}  
\figurenum{6c} 
\epsscale{0.9}
\plotone{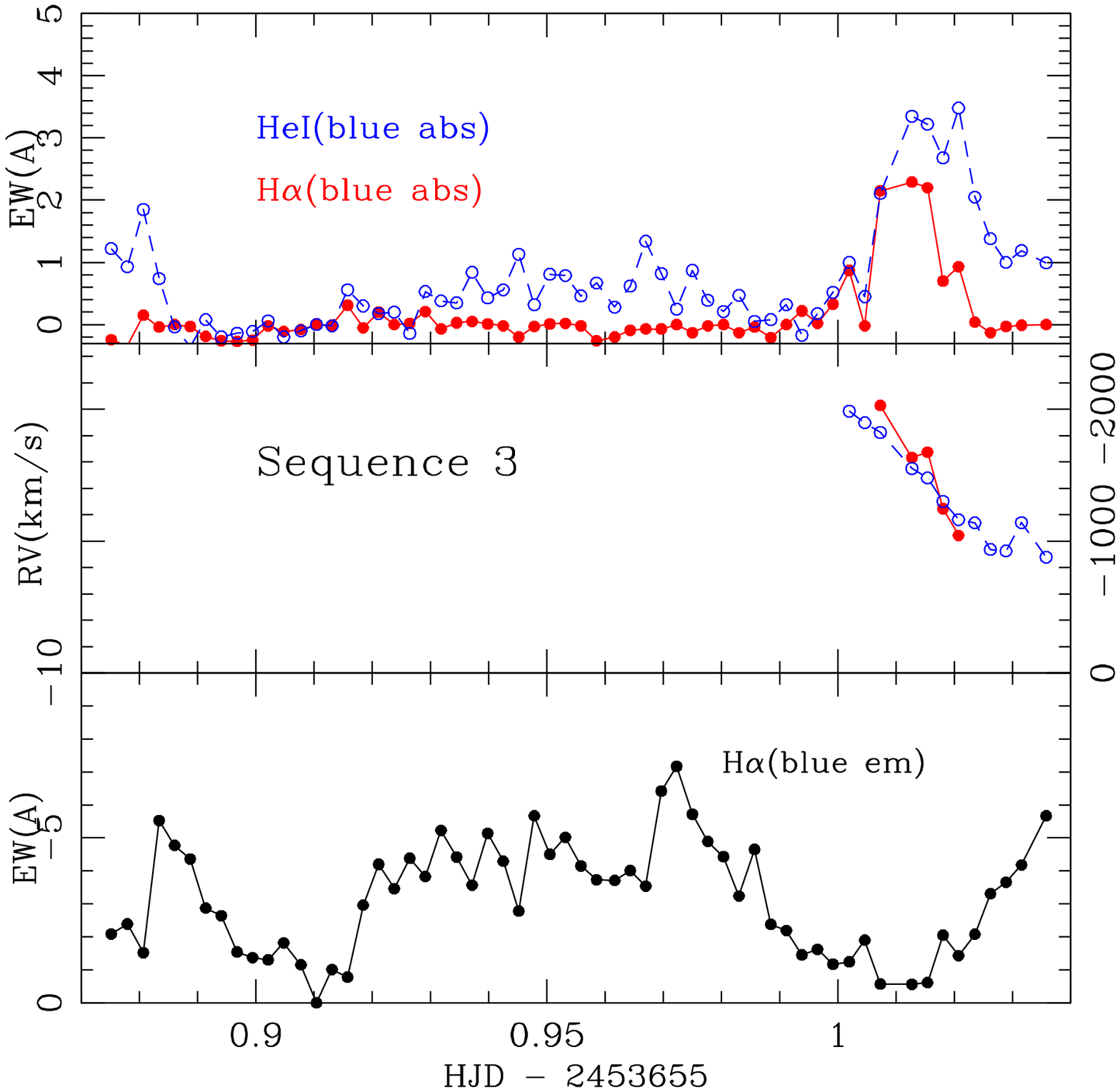}
\caption{Same as Figure 6a, except for spectral set S3.} 
\end{figure}
\clearpage  

\begin{figure}  
\figurenum{6d} 
\epsscale{0.9}
\plotone{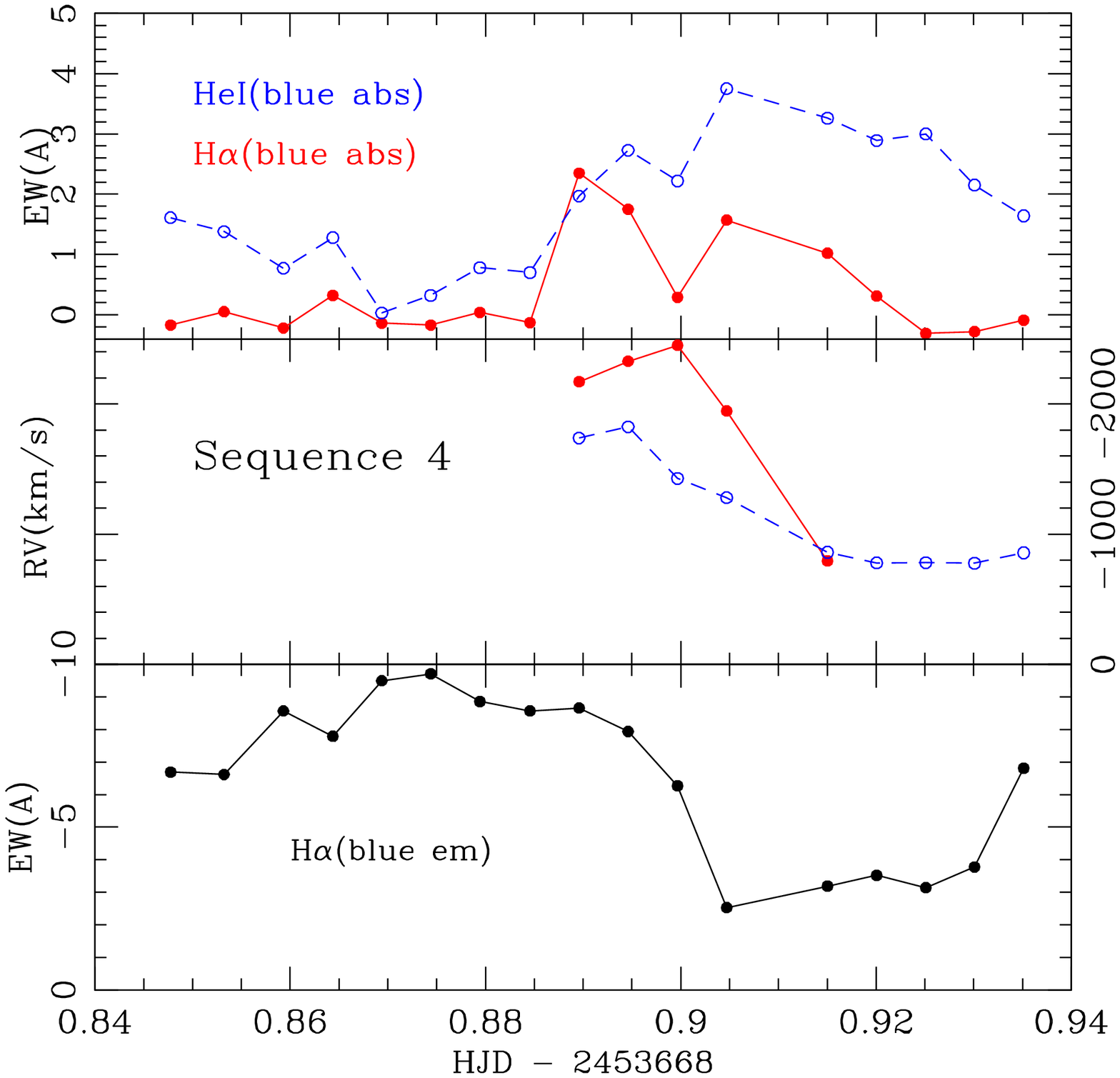}
\caption{Same as Figure 6a, except for spectral set S4.} 
\end{figure}
\clearpage  

\begin{figure}  
\figurenum{6e} 
\epsscale{0.9}
\plotone{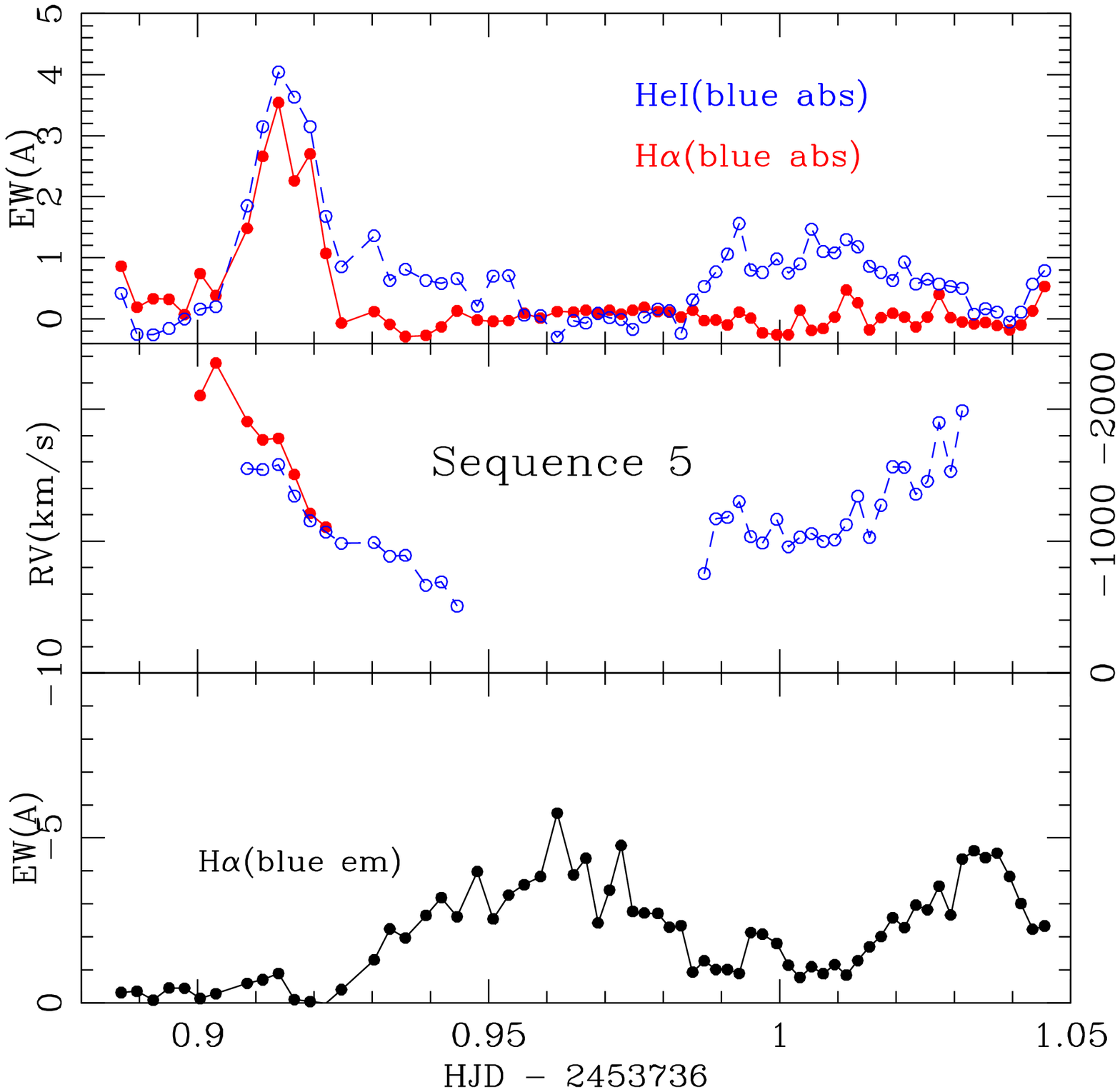}
\caption{Same as Figure 6a, except for spectral set S5.} 
\end{figure}
\clearpage  

\begin{figure}  
\figurenum{6f} 
\epsscale{0.9}
\plotone{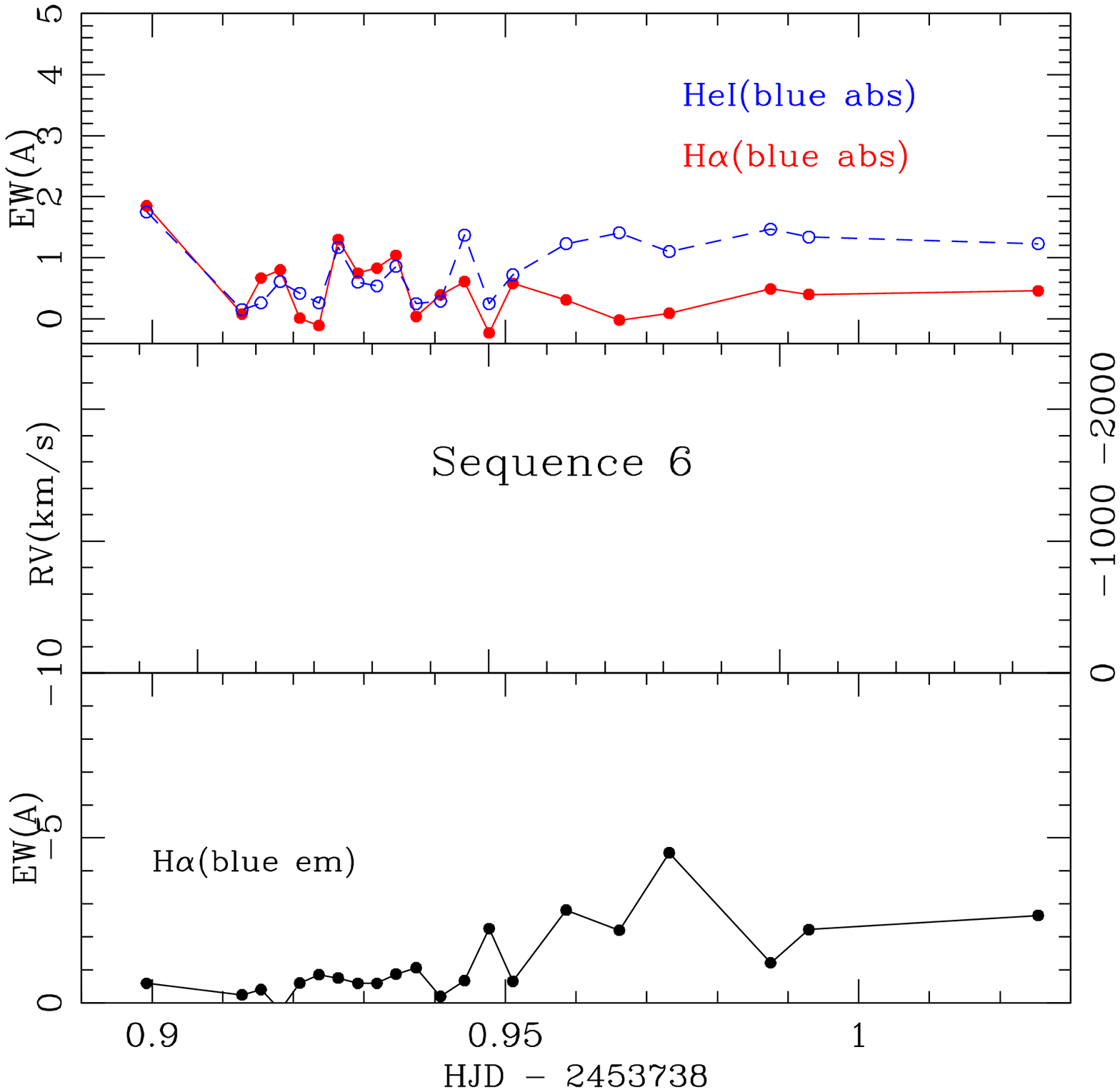}
\caption{Same as Figure 6a, except for spectral set S6.} 
\end{figure}
\clearpage  

\begin{figure}  
\figurenum{6g} 
\epsscale{0.9}
\plotone{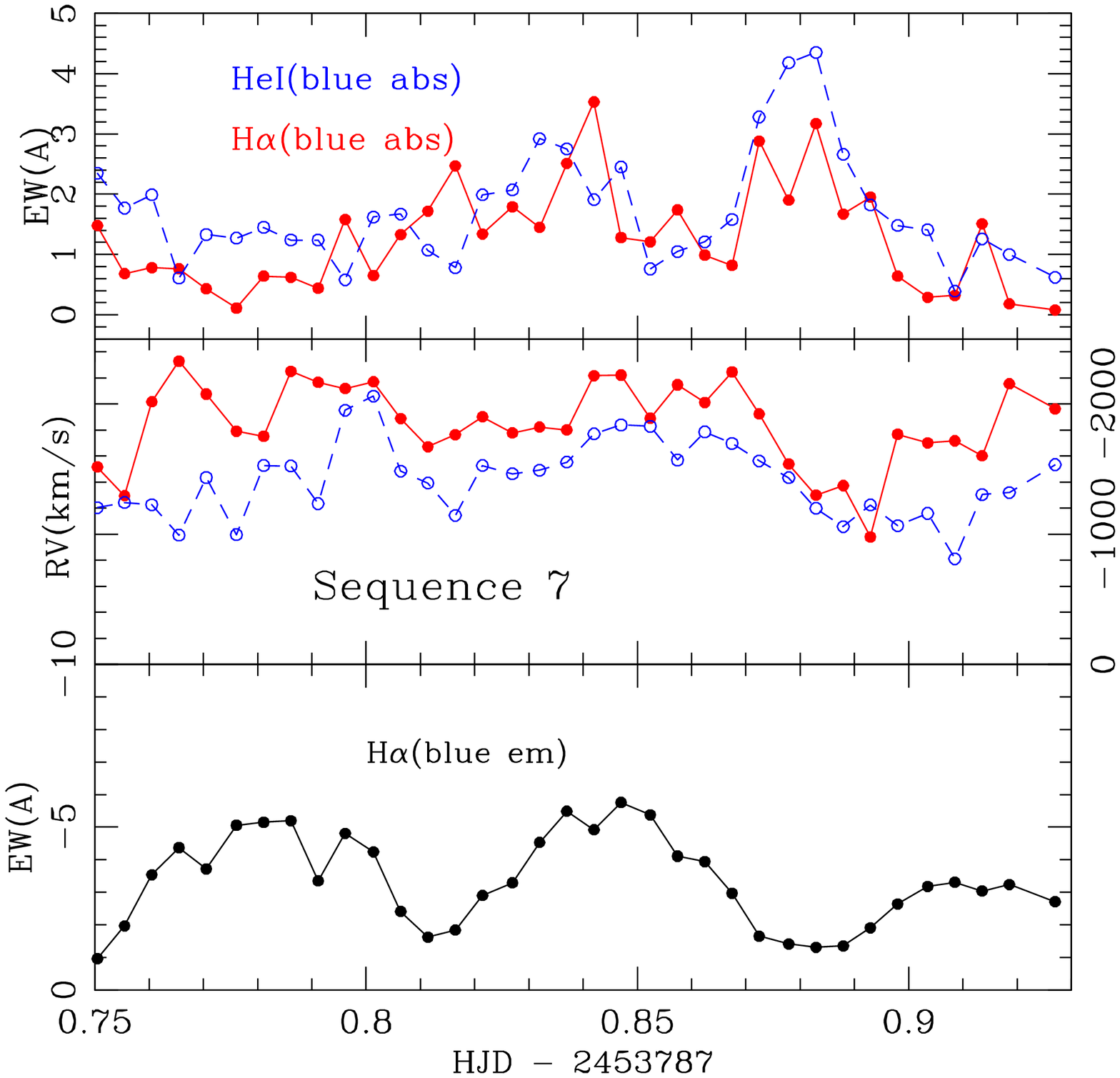}
\caption{Same as Figure 6a, except for spectral set S7.} 
\end{figure}
\clearpage  

\begin{figure}  
\figurenum{6h} 
\epsscale{0.9}
\plotone{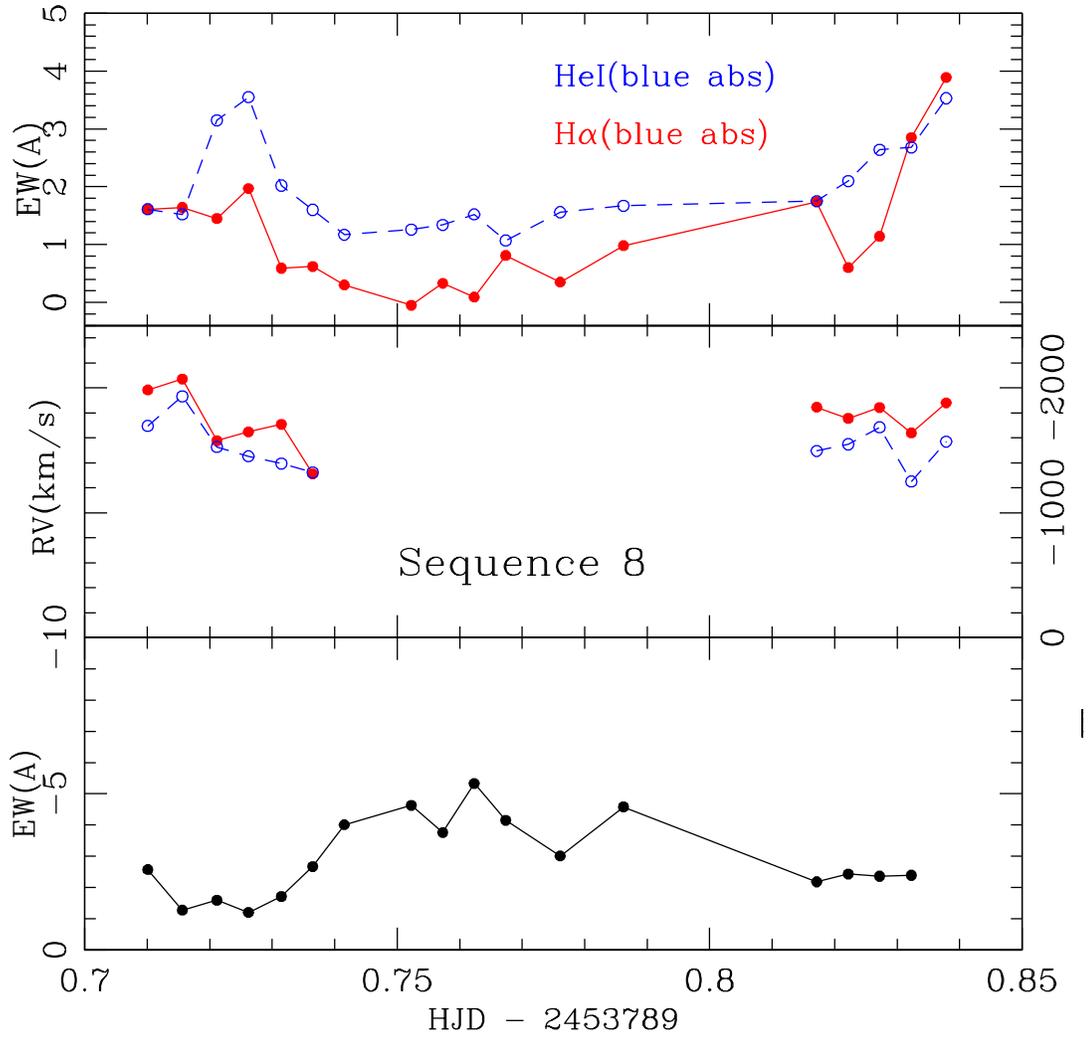}
\caption{Same as Figure 6a, except for spectral set S8.  Comment on
lack of data in middle panel.} 
\end{figure}
\clearpage  

\begin{figure}  
\figurenum{6i} 
\epsscale{0.9}
\plotone{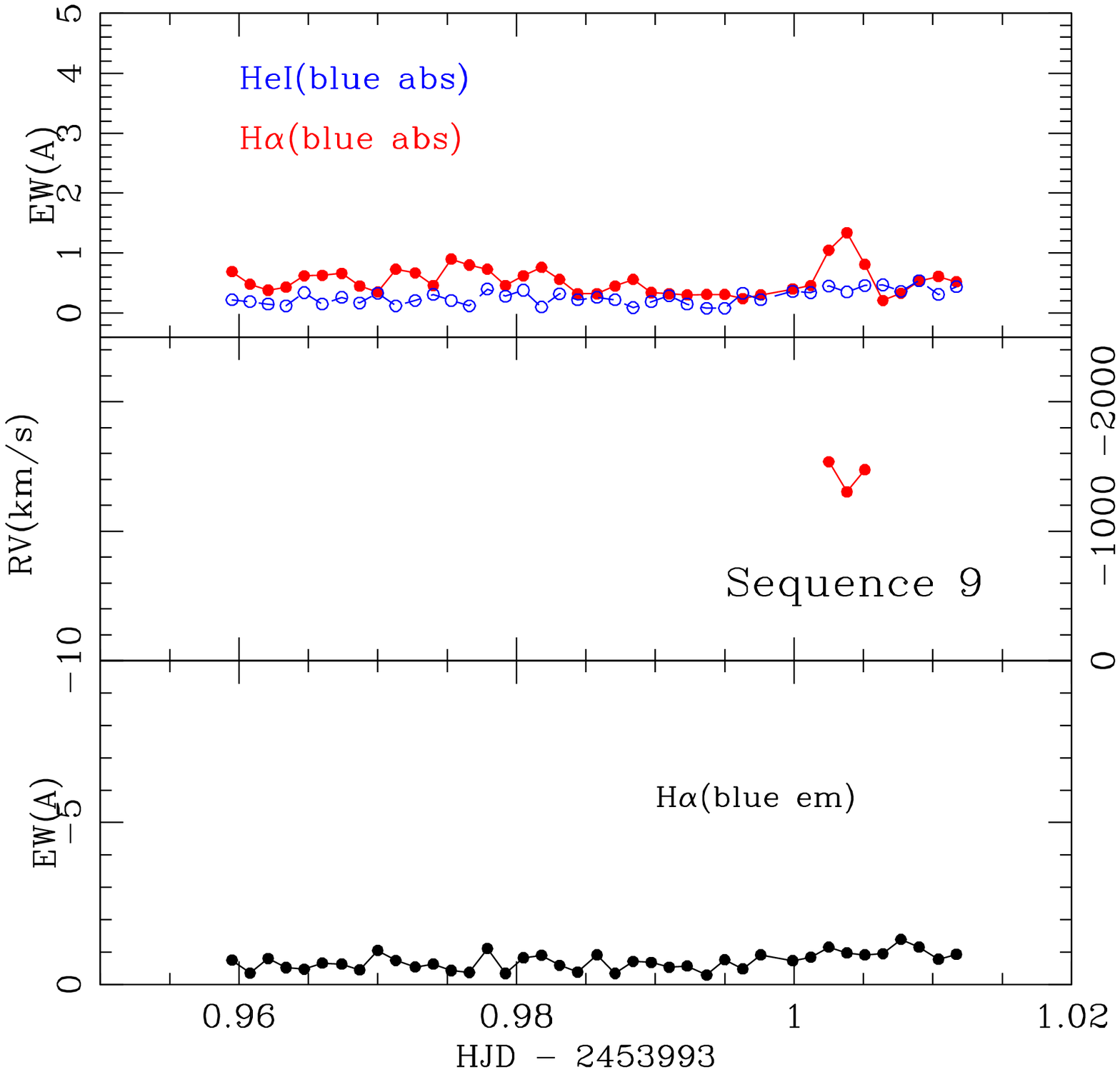}
\caption{Same as Figure 6a, except for spectral set S9.} 
\end{figure}
\clearpage  

\begin{figure}  
\figurenum{7} 
\epsscale{0.9}
\plotone{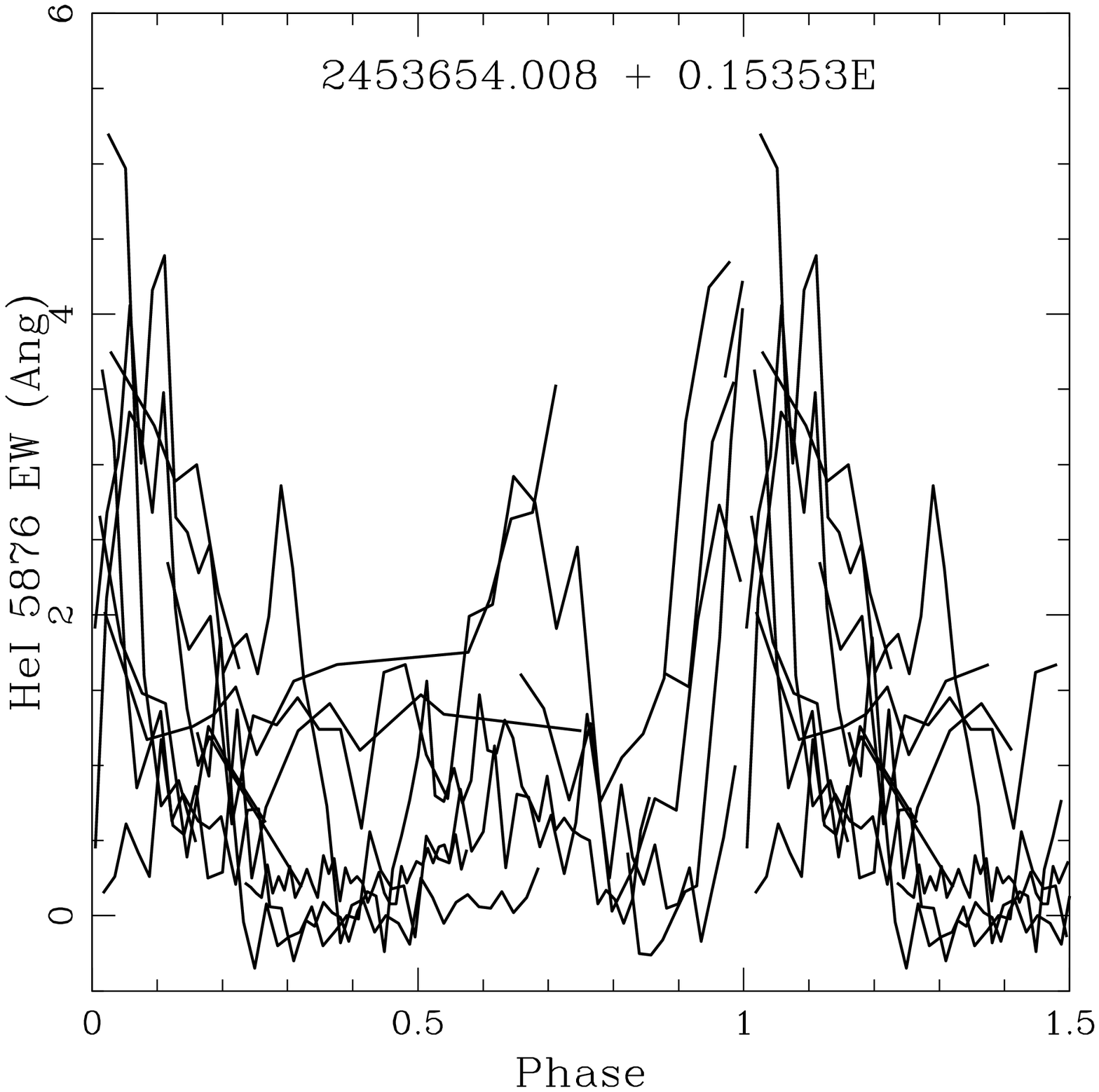}
\caption{The wind strength, as measured by the EW of the blueshifted
absorption in HeI 5576, vs orbital phase.  All of the EWs from a single
spectral set are connected by straight lines.  One can see both the
episodal nature of individual wind events as well as the concentration
of these event to orbital phases just after inferior conjuction of the
mass-gaining star.} 
\end{figure}
\clearpage  

\begin{figure}  
\figurenum{8} 
\epsscale{0.9}
\plotone{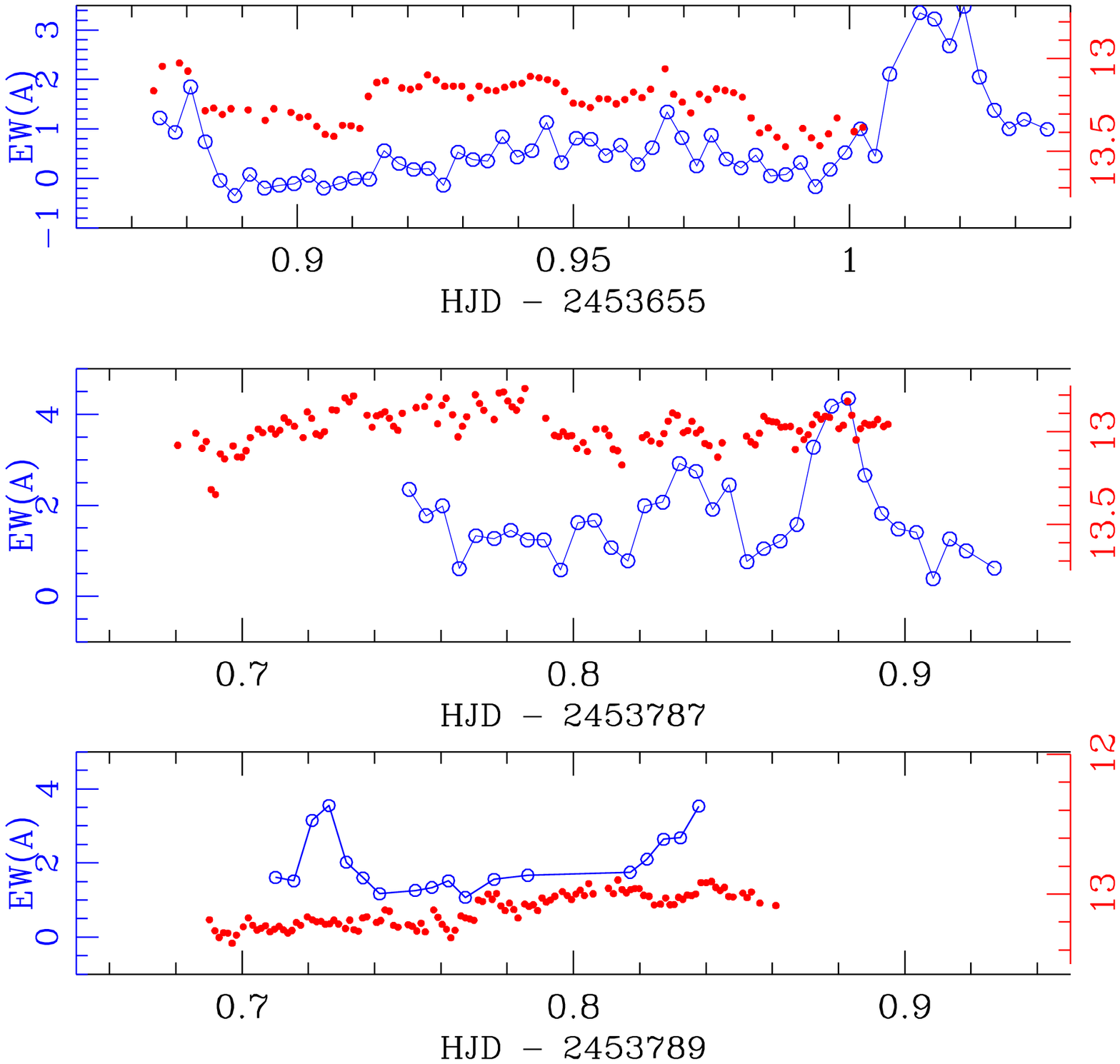}
\caption{Results from simultaneous spectroscopy and photometry
on three nights.  Top panel: Data from 2005-Oct-12 UT using photometric sequence 
P4 (closely-spaced small filled circles) and spectroscopic sequence S3 (larger
open circles connected by straight lines).  The spectroscopy measures the EW
of the blueshifted absorption in HeI 5876 (left axis) and the photometry is
V magnitude (right axis).  Middle panel:  Similar data from 2006-Feb-21 using
sequences P5 and S7.  Bottom panel:  Similar data from 2006-Feb-23 using sequences
P6 and S8.}
\end{figure}
\clearpage  

\end{document}